\documentclass[fleqn,usenatbib]{mnras}

\usepackage{newtxtext,newtxmath}

\usepackage[T1]{fontenc}

\DeclareRobustCommand{\VAN}[3]{#2}
\let\VANthebibliography\thebibliography
\def\thebibliography{\DeclareRobustCommand{\VAN}[3]{##3}\VANthebibliography}

\usepackage{graphicx}	
\usepackage{amsmath}	
\usepackage{siunitx} 
\usepackage[super]{nth}

\DeclareSIUnit\angstrom{\text {Å}}

\title[JWST/MIRI: stars and AGNs]{JWST/MIRI analysis of star formation and supermassive black hole growth from cosmic dawn to cosmic noon}

\author[J. C. J. D'Silva et al.]{
Jordan C. J. D'Silva,$^{1}$\thanks{E-mail: jordan.dsilva@manchester.ac.uk}
Christopher J. Conselice,$^{1}$
Duncan Austin,$^{1}$
Qiong Li,$^{1}$
Louis Quilley,$^{1}$ \newauthor
Vadim Rusakov,$^{1}$ 
Carys J. E. Gilbert,$^{1}$
Megan H. Martinez,$^{1}$
Julien Marabotto$^{1}$
\\
$^{1}$Jodrell Bank Centre for Astrophysics, Alan Turing Building, University of Manchester, Oxford Road, Manchester M13 9PL, UK
}

\date{Accepted XXX. Received YYY; in original form ZZZ}

\pubyear{\the\year{}}

\begin{document}
\label{firstpage}
\pagerange{\pageref{firstpage}--\pageref{lastpage}}
\maketitle

\begin{abstract}
We investigate the union of star formation and supermassive black hole (SMBH) growth from cosmic dawn to cosmic noon with the \textit{Mid-Infrared Instrument} (MIRI) on the \textit{James Webb Space Telescope} (\textit{JWST}). Photometric redshifts, stellar masses, star formation rates (SFRs) and active galactic nucleus (AGN) bolometric luminosities from a sample of $3218$ galaxies at $1 \lesssim z \lesssim 10$ are inferred by \textsc{EAZY} and \textsc{ProSpect} spectral energy distribution (SED) fitting. The fraction of galaxies hosting significant AGN ($\geq 10$ per cent of the fitted SED in the rest-frame UV or integrated $3-20 \, \mu \mathrm{m}$) across redshift and their AGN luminosities ($\mathrm{L_{AGN} \gtrsim 10^{44} \, erg \, s^{-1}}$) from broadband fitting are generally in agreement with spectroscopic results. Significant AGNs are especially prevalent in galaxies with $\mathrm{M_{\star} \lesssim 10^{9} \, M_{\odot}}$ and at $z \gtrsim 2$, and are located across the breadth of the $\mathrm{SFR-M_{\star}}$ plane. The ratio between the SFRs and the AGN matter accretion rates is on average $\lesssim 1$ in galaxies with $\mathrm{M_{\star} \lesssim 10^{7} \, M_{\odot}}$ at $z\gtrsim 3$. A toy model derived from simple accretion physics indicates that the ratio of their SMBH to stellar masses could have been as much as $\approx 3$~dex greater than local relations, facilitated by super-Eddington accretion (up to $\mathrm{\lambda_{Edd} \approx 3.7}$). This rapid growth of SMBHs and stars precedes the onset of star formation quenching and AGN downsizing. Lastly, alternative AGN SED models for a subsample of Little Red Dots (LRDs) were explored. Gaseous envelope models estimate a median of $\approx 0.6$~dex lower AGN luminosities than dusty torus models. The stellar masses of LRDs can be $\approx 2$~dex lower if the fitted Balmer break largely originates from a non-stellar component. 

\end{abstract}




\section{Introduction}
At present, the Universe is forming stars and growing supermassive black holes (SMBHs) at a rate $\approx 10$ times slower than at its peak at $z\approx 2$. Both the cosmic star formation history (CSFH) and cosmic black hole accretion rate history (CBHARH) are unimodal functions of redshift, characterised as rising swiftly after the onset of galaxy formation before turning over and steadily declining \citep{madauCosmicStarFormationHistory2014a}. The decay of both the CSFH and CBHARH is similar, indicative of a common depletion of the usable gas supply in galaxies from which stars are formed and SMBHs fed \citep{dsilvaGAMADEVILSCosmic2023a}. Correlations between the SMBH mass and stellar mass are observed in $z \lesssim 2$ galaxies \citep{kormendyCoevolutionNotSupermassive2013,greeneIntermediateMassBlackHoles2020a}, providing further evidence that these two baryon conversion processes have been coevolving for the last $\approx 10$~Gyr. 

How both net star formation and SMBH growth have matured since the onset of galaxy formation and up to the peak at $z\approx 2$ therefore sets up the conditions that led to their eventual, contemporaneous downfalls. The \textit{James Webb Space Telescope} (\textit{JWST}) has shown that UV luminous galaxies at $z\gtrsim 10$ are more abundant than previously expected, and the rapid assembly of stars shortly after the Big Bang is now well established in the canon of galaxy formation \citep[e.g.,][]{bouwensUVLuminosityDensity2023a,donnanEvolutionGalaxyUV2023a,harikaneComprehensiveStudyGalaxies2023,adamsEPOCHSIIUltraviolet2024,donnanJWSTPRIMERNew2024a,finkelsteinCompleteCEERSEarly2024a,robertsonEarliestGalaxiesJADES2024a,harikaneJWSTALMAKeck2025a,whitler9GalaxyUV2025,weibelExploringCosmicDawn2026}. 

At the same time, \textit{JWST} has shown that active galactic nuclei (AGNs) are similarly abundant at $z\gtrsim 4$, more so than would otherwise have been expected from pre-\textit{JWST} extrapolations of the AGN number density distributions \citep[e.g.,][]{juodzbalisEPOCHSVIIDiscovery2023,harikaneJWSTNIRSpecFirst2023,kocevskiHiddenLittleMonsters2023,greeneUNCOVERSpectroscopyConfirms2024,kokorevCensusPhotometricallySelected2024,mattheeLittleRedDots2024,kocevskiRiseFaintRed2025a,labbeUNCOVERCandidateRed2025,taylorBroadlineAGNs352025}. At $z\gtrsim 4$, SMBH growth through matter accretion outpaces star formation, with the ratio of SMBH mass to stellar mass in galaxies hosting an AGN exceeding local relations by $\gtrsim 1$~dex \citep{harikaneJWSTNIRSpecFirst2023,pacucciJWSTCEERSJADES2023,maiolinoJADESDiversePopulation2024}. 

Both the rapid assembly of stars and SMBHs are expected to inhibit gas conversion through feedback. Stars can affect gas condensation through photoionisation, which impedes gas cooling, and through both stellar winds and supernovae explosions, which can drive outflows in galaxies and lessen the fuel source for star formation \citep[e.g.,][]{hopkinsStellarFeedbackGalaxies2012}. Feedback from AGNs can impede the gas conversion efficiency in galaxies by expelling gas from the galaxies and heating up the gas in the circumgalactic medium \citep[CGM, e.g.,][]{beckmannCosmicEvolutionStellar2017,wrightBaryonCycleModern2024a,pacucciRedshiftEvolutionRelation2024}. Together, stellar and AGN feedback are important ingredients to reconcile the observed declines of the CSFH and CBHARH with predictions from galaxy formation models \citep{schayePhysicsDrivingCosmic2010,scharreEffectsStellarAGN2024}.

At $z \lesssim 10$, the CSFH reported in \textit{JWST} analyses tends to agree with those from earlier works from, for example, the \textit{Hubble Space Telescope} \citep[\textit{HST}, e.g.,][]{mcleod910GalaxyPopulation2016,oeschDearth10Galaxies2018,harikaneGOLDRUSHIVLuminosity2022}. This is consistent with a transition from star formation being rapid at $z\gtrsim 10$ to being largely dictated by structure formation and stellar/AGN feedback at later times \citep[e.g.,][]{harikaneEvolutionStellartoHaloMass2016,mitchellEvolutionStellarMass2016,rodriguez-pueblaConstrainingGalaxyhaloConnection2017,harikaneGOLDRUSHIIClustering2018a}. The balance between gas condensation and feedback leads to the tight positive correlation between stellar mass and star formation rates \citep[e.g.,][]{tacchellaConfinementStarformingGalaxies2016,elbazStarburstsOutStarformation2018a}, otherwise referred to as the star forming main sequence \citep[SFMS, e.g.,][]{popessoMainSequenceStarforming2023}. 

The SFMS has been shown to exist at $z\approx 9$, implying that self-regulation was underway in galaxies at this time \citep{simmondsBurstingSeamsStarforming2025}. Feedback also appears to be ubiquitous at $z\gtrsim 3$ as \textit{JWST} has detected populations of massive, $\mathrm{M_{\star} \gtrsim 10^{10} \, M_{\odot}}$, quenched galaxies \citep{carnallMassiveQuiescentGalaxy2023,glazebrookMassiveGalaxyThat2024,nanayakkaraPopulationFaintOld2024,nanayakkaraFormationHistoriesMassive2025,russellCosmicStillnessHigh2025a}. 


The \textit{Near-Infrared Camera} (NIRCam) instrument on \textit{JWST} has been monumental for measuring the distributions of stellar mass at $z\gtrsim 1$ \citep{weibelGalaxyBuildupFirst2024,shuntovCOSMOSWebStellarMass2025,harveyEPOCHSIVSED2025}. At $z\gtrsim 5$, NIRCam only covers $\lambda_{\mathrm{rest}} \lesssim 1 \, \mu \mathrm{m}$, which may result in overestimated stellar masses inferred by SED fitting \citep{songSolutionConflictEstimations2023,wangJWSTMIRIReveals2025}. In fact, the stellar mass density at $z\approx 9$ can be as much as $\approx 1$~dex overestimated without redder wavelength coverage than NIRCam \citep{papovichCEERSKeyPaper2023}. Observations at MIR observations, redder than NIRCam, are hence important to better measure the stellar masses and star formation rates of high redshift galaxies.

The \textit{Near-Infrared Spectrograph} (NIRSpec) has been critical for observing AGNs at $z\gtrsim 1$ \citep{kocevskiHiddenLittleMonsters2023,harikaneJWSTNIRSpecFirst2023,mattheeLittleRedDots2024,maiolinoJADESDiversePopulation2024}. Many of these AGNs at $z\gtrsim 4$ appear to be compact and red sources known as known as Little Red Dots \citep[LRDS, e.g.,][]{labbePopulationRedCandidate2023, greeneUNCOVERSpectroscopyConfirms2024, mattheeLittleRedDots2024, kokorevCensusPhotometricallySelected2024, kocevskiRiseFaintRed2025a} with striking `V-shaped' spectral energy distributions (SEDs). Spectroscopically targeted AGNs may not, however, be representative of the prevailing population implying that conclusions drawn on the growth of SMBHs at high redshift could be subject to selection effects \citep{liTipIcebergOvermassive2025,brooksMonstersMoreComplete2026}. While NIRCam observations can mitigate some of these selection effects, it has been argued that only lower and upper limits could be placed on the AGN contribution to galaxy SEDs owing to the difficulty in separating AGN from host emission with only the rest-frame UV-optical \citep{dsilvaSelfconsistentJWSTCensus2025a}. 

The \textit{Mid-Infrared Instrument} (MIRI) onboard the \textit{JWST} is the redder companion to NIRCam, extending the observable wavelengths of the telescope up to $\mathrm{\lambda_{obs} \approx 25\, \mu m}$. This has so far enabled the measurement of obscured AGNs up to $z\approx 8$ and accurate constraints on stellar masses and star formation rates, improving on similar results obtained from NIRCam alone \citep{yangCEERSKeyPaper2023,lyuActiveGalacticNuclei2024a,liEPOCHSPaperVIII2024,carranza-escuderoLonelyLittleRed2025}. 

Rapid stellar mass assembly, precocious growth of SMBHs and the onset of quenching motivate us to explore the union of star formation and AGN activity in galaxies from cosmic noon to cosmic dawn. In this paper, we appeal to multiwavelength photometric observations, especially with MIRI, and leverage SED fitting to study the union of stellar mass assembly and SMBH growth. An important aspect of this work is that we aim to account for stars and AGNs simultaneously when modelling the SEDs. 

The structure of this paper is outlined as follows. The MIRI data and photometric extraction are described in Section~\ref{sec:data}. Our methods of SED fitting and AGN identification are detailed in Section~\ref{sec:sedfitting}. Results and discussions on the demographics of AGNs are presented in Section~\ref{sec:agndemographics}. Our conclusions are listed in Section~\ref{sec:conclusions}. We use the AB magnitude system throughout \citep{okeSecondaryStandardStars1983a}. We use standard, `concordance' $\Lambda$CDM cosmology with $H_{0} = 70 \, \mathrm{km s^{-1}\, Mpc^{-1}}$, $\Omega_{\Lambda} = 0.7$ and $\Omega_{\mathrm{M}} = 0.3$. Throughout, we denote the normal distribution with mean, $\mu$, and variance, $\sigma^{2}$, as $\mathcal{N}(\mu, \sigma^{2})$ and the uniform distribution between bounds $a$ and $b$ as $U(a,b)$. 

\section{Data} \label{sec:data}
In this paper, we use \textit{JWST} MIRI imaging obtained from the Systematic Mid-infrared Instrument Legacy Extragalactic Survey \citep[SMILES,][]{riekeSMILESPrototypeJWST2024a, albertsSMILESInitialData2024a} and the MIRI Deep Imaging Survey \citep[MIDIS,][]{ostlinMIRIDeepImaging2025a}. 

SMILES includes F560W, F770W, F1000W, F1280W, F1500W, F1800W, F2100W and F2550W imaging, covering $\approx 35 \, \mathrm{arcmin^{2}}$ in the southern pointing of the Great Observatories Origins Deep Survey (GOODS-S) field \citep{giavaliscoGreatObservatoriesOrigins2004}. MIDIS includes the F560W and F770W bands, covering $\approx 5.5 \, \mathrm{arcmin^{2}}$ in the Hubble Ultra Deep Field \citep[HUDF,][]{williamsHubbleDeepField1996,beckwithHubbleUltraDeep2006,illingworthHSTEXtremeDeep2013}. SMILES mosaics were downloaded from the Hubble Legacy Science Product data release\footnote{\url{https://archive.stsci.edu/hlsp/smiles}} and the MIDIS mosaics were downloaded from the MIDIS data release \citep{melinder_2025_15624625}. We highlight that both of these datasets were processed using the \texttt{JWST Calibration Pipeline} \citep[e.g.,][]{bushouse_2023_10022973} and included custom background removal steps \citep[e.g.,][]{perez-gonzalezWhatNatureLittle2024a,jermannMIDISIdentificationDeep2026}.

MIDIS is circumscribed by SMILES in the GOODS-S field meaning that $\mathrm{\lambda_{obs} \gtrsim 7.6 \, \mu m}$ photometric points/upper limits could be obtained for sources only present in MIDIS and not SMILES. MIDIS reaches $\approx 2 \, \mathrm{mag}$ deeper in the overlapping filters than SMILES. On the other hand, by surveying $\approx 6$ times more area, SMILES is important for measuring the light from brighter sources that occupy larger volumes.

Both the GOODS-S field and HUDF are complemented by extensive NIRCam observations from \textit{JWST}, covering $\mathrm{\lambda_{obs} \approx 0.9-4.8 \, \mu m}$ imaging, and \textit{HST} observations, especially with the Advanced Camera for Surveys instrument (ACS) that covers $\mathrm{\lambda_{obs} \approx 0.4-0.9 \, \mu m}$. We folded in these ancillary data from the EPOCHS dataset (\citealt{adamsEPOCHSIIUltraviolet2024,conseliceEPOCHSDiscoveryStarforming2025, austinEPOCHS2026}). The EPOCHS NIRCam data in the fields overlapping with the relevant MIRI observations comprise of images from JADES \citep[e.g.,][]{eisensteinOverviewJWSTAdvanced2026a} and NGDEEP \citep[e.g.,][]{bagleyNextGenerationDeep2024b}. The ACS data was obtained from the Hubble Legacy Fields data release in the GOODS-S field \citep[e.g.,][]{illingworthHubbleLegacyFields2016,whitakerHubbleLegacyField2019}. The depths and effective areas of all the data are summarised in Table~\ref{tab:field_summary}.

\begin{table*}
    \begin{tabular}{c|c|c|c|c}
         & SMILES & MIDIS & SMILES & MIDIS \\
         Filter & $5\sigma$ depth $[\mathrm{mag}]$ & $5\sigma$ depth $[\mathrm{mag}]$ & Area $[\mathrm{arcmin^{2}}]$ & Area $[\mathrm{arcmin^{2}}]$ \\
         \hline
        \textit{HST}/ACS F435W & 28.06 & 29.16 & 34.46 & 5.47 \\
        \textit{HST}/ACS F606W & 28.40 & 29.44 & 34.46 & 5.47 \\
        \textit{JWST}/NIRCam F070W & 27.80 & -- & 0.72 & -- \\
        \textit{HST}/ACS F775W & 27.84 & 29.16 & 34.46 & 5.47 \\
        \textit{HST}/ACS F814W & 28.01 & 28.05 & 34.46 & 5.47 \\
        \textit{JWST}/NIRCam F090W & 28.74 & 29.00 & 34.23 & 5.47 \\
        \textit{HST}/ACS F850LP & 27.44 & 28.48 & 34.46 & 5.47 \\
        \textit{JWST}/NIRCam F115W & 28.71 & 29.02 & 34.23 & 5.47 \\
        \textit{JWST}/NIRCam F150W & 28.65 & 29.02 & 34.23 & 5.47 \\
        \textit{JWST}/NIRCam F162M & 28.51 & -- & 2.45 & -- \\
        \textit{JWST}/NIRCam F182M & 28.40 & 28.55 & 13.11 & 4.46 \\
        \textit{JWST}/NIRCam F200W & 28.64 & 29.04 & 34.23 & 5.47 \\
        \textit{JWST}/NIRCam F210M & 28.33 & 28.43 & 13.05 & 4.44 \\
        \textit{JWST}/NIRCam F250M & 28.50 & -- & 2.39 & -- \\
        \textit{JWST}/NIRCam F277W & 28.84 & 29.19 & 34.27 & 5.47 \\
        \textit{JWST}/NIRCam F300M & 28.72 & -- & 2.36 & -- \\
        \textit{JWST}/NIRCam F335M & 28.75 & 28.96 & 28.99 & 5.47 \\
        \textit{JWST}/NIRCam F356W & 28.91 & 29.25 & 34.27 & 5.47 \\
        \textit{JWST}/NIRCam F410M & 28.75 & 29.09 & 34.27 & 5.47 \\
        \textit{JWST}/NIRCam F430M & 27.92 & 27.99 & 10.29 & 4.34 \\
        \textit{JWST}/NIRCam F444W & 28.94 & 29.26 & 34.27 & 5.47 \\
        \textit{JWST}/NIRCam F460M & 27.47 & 27.54 & 10.22 & 4.34 \\
        \textit{JWST}/NIRCam F480M & 27.46 & 27.47 & 10.17 & 4.30 \\
        \textit{JWST}/MIRI F560W & 25.97 & 28.08 & 34.46 & 4.50 \\
        \textit{JWST}/MIRI F770W & 25.95 & 27.95 & 34.11 & 4.60 \\
        \textit{JWST}/MIRI F1000W & 25.06 & 25.16* & 34.02 & 5.42 \\
        \textit{JWST}/MIRI F1280W & 24.72 & 24.75* & 34.11 & 5.46 \\
        \textit{JWST}/MIRI F1500W & 24.66 & 24.71* & 34.11 & 5.46 \\
        \textit{JWST}/MIRI F1800W & 23.89 & 23.89* & 34.11 & 5.46 \\
        \textit{JWST}/MIRI F2100W & 23.64 & 23.63* & 34.10 & 5.46 \\
        \textit{JWST}/MIRI F2550W & 21.86 & 21.89* & 34.12 & 5.46 \\
        \hline
    \end{tabular}
    \caption{Depths and effective areas of the SMILES and MIDIS data used in this work in wavelength order. 5$\sigma$ sensitivity limits are obtained from random sampling in blank regions with $0.\arcsec32$ radius apertures. The effective area is only of the unmasked regions. MIDIS depths marked by an asterisk symbol are effectively from SMILES data in the overlapping region (see text for details).}
    \label{tab:field_summary}
\end{table*}

\subsection{Photometric selection and extraction} \label{sec:photo_extraction}
Photometric measurements from the MIRI, NIRCam and ACS images were extracted using the source-finding code \textsc{ProFound} \citep{robothamProFoundSourceExtraction2018}. A key feature of \textsc{ProFound} is its implementation of source deblending where segmentation maps conform to the natural morphologies of sources, including light scattered by the point spread function (PSF), and curve-of-growth dilations to capture total fluxes. \textsc{ProFound} has been used extensively for source-extraction in the past for both ground-based surveys \citep[e.g.,][]{bellstedtGalaxyMassAssembly2020a,daviesDeepExtragalacticVIsible2021b} and space-based surveys with \textit{JWST} and \textit{HST} \citep[e.g.,][]{dsilvaSelfconsistentJWSTCensus2025a,tompkinsSKYSURFIXCosmic2026b}. 

\begin{figure*}
    \centering
    \includegraphics[width=\linewidth]{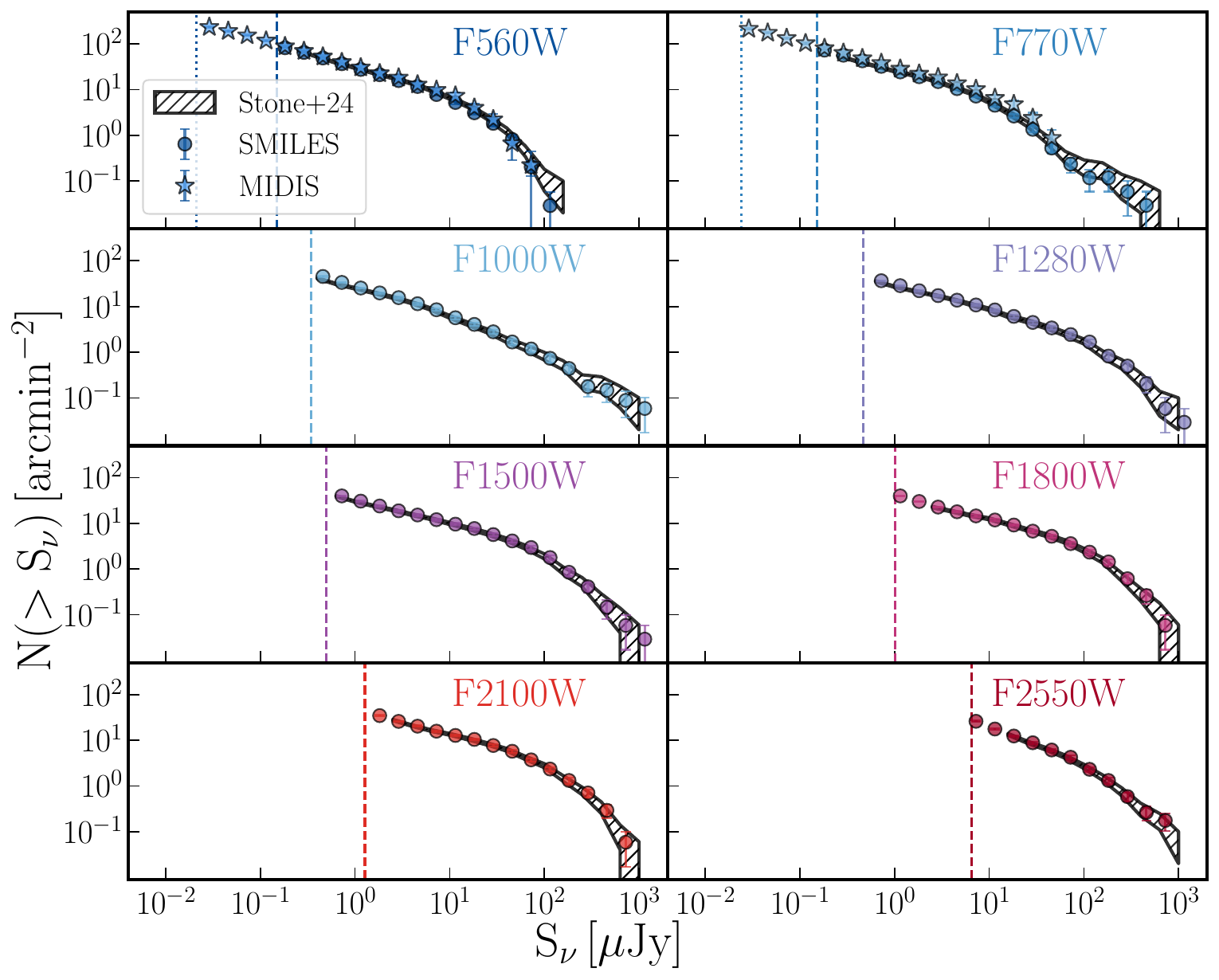}
    \caption{Cumulative number counts in 8 MIRI bands as indicated in the figure legend. Error bars denote Poisson uncertainties in each bin. The dotted/dashed vertical lines are the 5$\sigma$ sensitivity limits of MIDIS/SMILES in each band, obtained from random sampling in blank regions with $0.\arcsec32$ apertures. The black hatched curve shows the results and $1\sigma$ uncertainties from \citet{stone525MmGalaxy2024}. MIDIS observations are only in the F560W and F770W bands.}
    \label{fig:number_counts}
\end{figure*}

Initial source finding was applied to an inverse-variance weighted stack of the F560W and F770W images. Because the images are north-aligned and not at the native position angle of the telescope at the epoch of observation, pixels outside the observing footprint are set to \texttt{NA} and masked. We also masked pixels within $7$ pixels of the image edge to prevent low $S/N$ artefacts from entering our sample. We queried the GAIA catalogue \citep{vallenariGaiaDataRelease2023} and placed circular masks on bright stars in the fields, the radii of which were calculated from curve-of-growth measurements to cover the length of the diffraction spikes. The segmentation map from the F560W$+$F770W stack was then placed on the remaining images to extract forced-aperture photometry. Importantly, \textsc{ProFound} measures undilated, `colour' fluxes, where the segmentation map is tightly confined around the sources, and dilated, `total' fluxes where the segmentation map is allowed to grow until the fractional change in the added light does not exceed $5$ per cent. Colour fluxes are hence useful for high $S/N$ constraints of spectral features for especially photometric redshifts and total fluxes are useful for constraints of the net luminosity, and hence the total stellar mass in galaxies. 

We experimented with fluxes measured in circular apertures of $0.\arcsec 16 - 0.\arcsec 60$ radii but preferred to use the colour and total fluxes to avoid having to rely on band dependent aperture corrections. Depths were calculated from a random sampling of apertures placed in source-masked, blank parts of the images. Similarly, photometric uncertainties were computed from the distribution of sky pixels of the 200 nearest random apertures nearest to the given source \citep[e.g.,][]{adamsEPOCHSIIUltraviolet2024}. Flux uncertainties from these local depth measurements are $\approx 2-3$ times larger than those extracted from the ERR extensions produced by the standard pipeline due to correlated noise \citep{albertsSMILESInitialData2024a}. To account for systematic variations between bands, zero-point offsets for example, we applied a $10$ per cent uncertainty floor, added in quadrature, on all photometric measurements.

Small sources with half-light radius from \textsc{ProFound} approximately less than the PSF size in the detection bands, $S/N > 10$ and $\mathrm{|F560W-F770W|>0.5}$ were flagged as potentially hot pixels. In this work, we prioritise sources with the most complete spectral coverage, necessary to constrain astrophysical quantities, and so we required that every galaxy had a match against existing objects in the EPOCHS NIRCam catalogue. This could mean that we miss NIRCam dark galaxies, which may correspond to ultra-high redshift ($z \approx 17-25$) sources \citep[e.g.,][]{perez-gonzalezNIRCamdarkGalaxyDetected2024,perez-gonzalezRiseGalacticEmpire2025}. These galaxies are not expected to be detected in the SMILES data given the depths \citep{albertsSMILESInitialData2024a} and are expected to be rare given the survey volume of MIDIS \citep[e.g.,][]{jermannMIDISIdentificationDeep2026}. Out of 6964 detected sources 2267 were flagged as spurious, leaving us with a total of 4697 detected objects. 

In the best case scenario where all the images overlap on the sky we have 31-band photometry covering $\mathrm{\lambda_{obs} \approx  0.4 - 25.4\, \mu m}$. However, not every band overlaps with the entire SMILES/MIDIS footprint, as indicated by the effective areas in Table~\ref{tab:field_summary}. There were only 72 sources in the F070W, $\approx 200$ sources in the F162M, F250M, F300M and $\approx 2000$ sources in the F182M, F210M, F430M, F480M. For sources detected in both SMILES and MIDIS, we preferred to keep the measurement from SMILES to maintain uniformity with the other MIRI bands. This means that MIDIS was only used for the faintest sources. 

We cross matched our MIRI selected catalogue against the brown dwarf star catalogues in GOODS-South from \citet{hainlineJADESAbundanceUltradistant2026a}. A total of only three sources were found and subsequently removed from the extragalactic catalogue in the remaining analysis. Given the location of the SMILES/MIDIS field in relation to the galactic disk and the area covered by the observations, we expect that brown dwarf contamination is negligible \citep{hainlineBrownDwarfCandidates2024,tuThreeBrownDwarfs2025}.

Figure~\ref{fig:number_counts} shows the cumulative number counts per unit area of the 8 MIRI bands studied in this work. From this figure, it is clear the different regimes that SMILES and MIDIS occupy as mentioned previously. We compare this to the number counts from \citet{stone525MmGalaxy2024} who perform a similar analysis on the SMILES data with \textsc{PHOTUTILS} \citep{larry_bradley_2023_1035865}, where we find that our results are in agreement across the breadth of the flux distribution, assuring us that our photometric extraction and use of \textsc{ProFound} total fluxes is adequate. Note that only the unmasked area is included in the number counts.

\section{SED Fitting} \label{sec:sedfitting}
\subsection{Photometric redshifts} \label{sec:photoz}
The photometric redshift fitting code \textsc{EAZY} was used with the `\texttt{agn\_sfhz\_blue}' template set. This includes redshift dependent star formation histories (SFHs) to disfavour physically unrealistic episodes of star formation before the Big Bang. The template set is also complemented by the NIRSpec spectrum of an extreme emission line galaxy at $z_{\mathrm{spec}} = 8.5$ \citep{carnallFirstLookSMACS07232023} and a spectrum generated to match the NIRSpec spectrum of a LRD at $z_{\mathrm{spec}} = 4.53$ \citep{killiDecipheringJWSTSpectrum2024}, with a red optical continuum that is similar to a dust-reddened AGN torus. This template set was previously found to result in the tightest $z_{\mathrm{spec}} - z_{\mathrm{photoz}}$ constraints when compared to other EAZY template sets \citep{austinEPOCHS2026}.

In total there are 15  template spectra that \textsc{EAZY} can combine linearly to fit the photometric redshift and infer the full posterior distribution, $P(z_{\mathrm{phot}})$. The photometric redshift was taken to be at the maximum likelihood solution ($z_{\mathrm{ML}}$) from the template fitting and the uncertainty, $\sigma_{z_{\mathrm{ML}}}$, was obtained from the Laplace approximation and the local curvature of the $P(z_{\mathrm{phot}})$ about $z_{\mathrm{ML}}$. 

\begin{figure*}
    \centering
    \includegraphics[width=\textwidth]{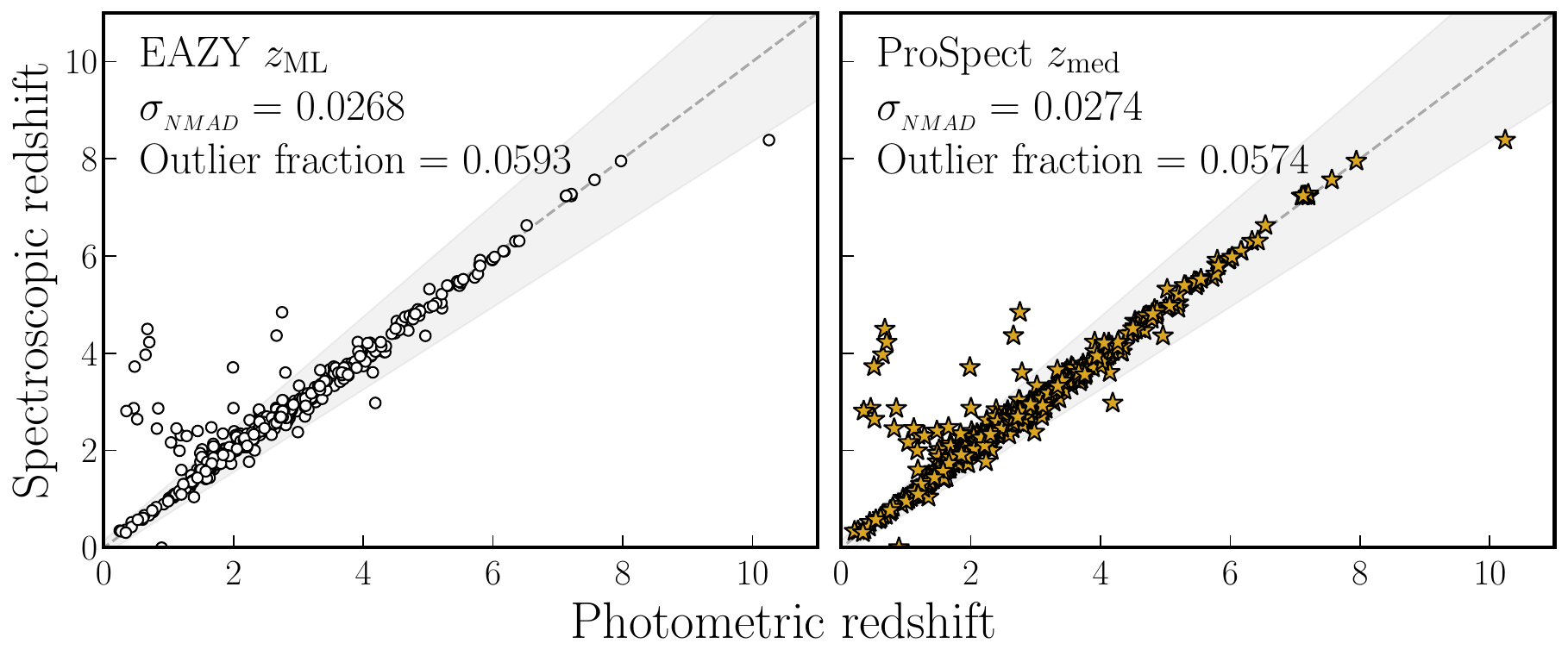}
    \caption{\textit{Left:} spectroscopic redshift against photometric redshift obtained with \textsc{EAZY}. \textit{Right:} same as the left panel except comparing to the photometric redshift obtained from \textsc{Prospect} fitting. The $\sigma_{\mathrm{NMAD}}$ and outlier fraction are reported in the panels, and are obtained from Equations~\ref{eq:sigmanmad} and \ref{eq:outlierfrac}. The dashed line in every panel is the one-to-one line. The grey filled region is the 15 per cent outlier contour.}
    \label{fig:zphot_zspec}
\end{figure*}

Figure~\ref{fig:zphot_zspec} shows a comparison between our inferred $z_{\mathrm{phot}}$ and 540 NIRSpec $z_{\mathrm{spec}}$ from the JADES spectroscopic sample in the GOODS-S field \citep{curtis-lakeJADESDataRelease2026,scholtzJADESDataRelease2026}. Denoting $\Delta z = z_{\mathrm{phot}}- z_{\mathrm{spec}}$, the performance of the photometric redshift is measured using the normalised-absolute-difference (NMAD), 
\begin{equation}\label{eq:sigmanmad}
    \sigma_{\mathrm{NMAD}} = 1.48\times \mathrm{med} \left( \frac{|\Delta z - \mathrm{med}(\Delta z)|}{1 + z_{\mathrm{spec}}} \right),
\end{equation}
and the outlier fraction, where an outlier is defined as a galaxy where 
\begin{equation}\label{eq:outlierfrac}
    \frac{|\Delta z|}{1+z_{\mathrm{spec}}} > 0.15.
\end{equation}
It can be seen from Figure~\ref{fig:zphot_zspec} that the photometric redshift recovers the correct redshift in most cases, with $\approx 5.9$ per cent outliers, which is similar to other analyses \citep[e.g.,][]{shuntovCOSMOS2025COSMOSWebGalaxy2025}. Most of the outliers are clustered at $z_{\mathrm{phot}} \lesssim 1$. The $z_{\mathrm{phot}}$ agrees well with the $z_{\mathrm{spec}}$ at $z\gtrsim1$ (and especially $z\gtrsim4$) where the main spectral feature to be fit is the Lyman break\footnote{Of course, degeneracies between the Lyman break and Balmer break can still cause $z_{\mathrm{phot}}$ errors.}. We choose to concentrate on galaxies at $1 \lesssim z_{\mathrm{phot}} \lesssim 10$ to avoid obvious outliers at $z < 1$ and to cover cosmic noon to cosmic dawn. 

\subsection{\textsc{ProSpect} fitting} \label{sec:prospect}
Fully Bayesian and generative SED fitting allows us to obtain meaningful astrophysical quantities such as stellar mass, star formation rate (SFR) and potential AGN contribution, and their propagated uncertainties. We used the SED fitting code \textsc{ProSpect} \citep{robothamProSpectGeneratingSpectral2020} and added the \textsc{EAZY} results as a prior of the form $\mathcal{N}(z_{\mathrm{ML}}, \sigma_{z_{\mathrm{ML}}}^{2})$. 

\textsc{ProSpect} models star light with the \citet{bruzualStellarPopulationSynthesis2003} stellar population synthesis library and the \citet{chabrierGalacticStellarSubstellar2003} initial mass function (IMF). The SFH is modelled with a 4 parameter smooth function similar to a skewed log-normal function, referred to as a \texttt{mass\_func\_snorm\_trunc} function \citep[see Eqs. 1-5 in][]{dsilvaGAMADEVILSCosmic2023a}. The SFH is flexible enough to permit a range of SFH shapes, from rising/star forming and declining/quenching SFHs \citep{daviesDeepExtragalacticVIsible2025a,daviesDeepExtragalacticVIsible2025b}. This SFH implementation has also been shown to have adequate flexibility to model the average shapes of SFHs from galaxy formation simulations \citep{lagosSharkIntroducingOpen2018a,robothamProSpectGeneratingSpectral2020}. \textsc{ProSpect} also models the metallicity as an evolving quantity that scales broadly with the assembly of stellar mass and chemical enrichment of the interstellar medium \citep[ISM, ][]{thorneDEVILSCosmicEvolution2022a}. This metallicity history combined with the flexible \texttt{mass\_func\_snorm\_trunc} SFH has been shown to be vital for the reconstruction of the cosmic star formation history through both $z=0$ SFH stacking experiments \citep{bellstedtGalaxyMassAssembly2020b} and synthesis models of the extragalactic background light (\citealt{koushanGAMADEVILSConstraining2021}; Driver et al. submitted).

Emission lines are modelled using the line tables from \citet{levesqueTheoreticalModelingStarForming2010}, which itself originate from the \textsc{mappings-III} photoionisation code \citep{sutherlandMAPPINGSIIIModelling2013}, and the strength of the radiation field is tied to the metallicity through the models of \citet{orsiNebularEmissionStarforming2014}. Absorption from neutral hydrogen in the intergalactic medium is modelled using the analytic model from \citet{inoueUpdatedAnalyticModel2014a} that accounts for the statistical distribution of the Lyman alpha forest, Lyman limit systems and damped Lyman alpha systems. 

Dust attenuation is implemented with the two-component \citet{charlotSimpleModelAbsorption2000b} model. The first component describes a prevailing screen of dust in the ISM that affects all sources of light, i.e., both stars and AGN. The second component accounts for dust in stellar birth clouds and only affects young stars with ages $\lesssim 10^{7} \, \mathrm{yr}$. We impose a prior on this model by assuming that the V-band optical depth must be greater in the birth clouds than in the comparatively more diffuse ISM. The absorbed energy is assumed to be re-radiated according to the \citet{daleTwoparameterModelInfrared2014a} dust emission models. We impose a normal prior on effectively the dust temperature through the $\alpha$ parameters, $\alpha \sim \mathcal{N}(2, 1^{2})$, in the \cite{daleTwoparameterModelInfrared2014a} models because this is largely unconstrained with our photometry that only goes as redward as $\lambda_{\mathrm{obs}} \approx 25.4\, \mu \mathrm{m}$ \citep{dsilvaContributionStarsDust2026a}.

The fiducial AGN component is implemented using the dusty torus models of \citet{fritzRevisitingInfraredSpectra2006} and later expanded on by \citet{feltreSmoothClumpyDust2012}. We fix some parameters that control the geometry and covering factor of the torus, to be consistent with the set up used in \citet{thorneDeepExtragalacticVIsible2022}. The key parameters for the AGN that we fit are the bolometric luminosity of the AGN, $\mathrm{L_{AGN}}$, the viewing angle and the optical depth at $9.7 \, \mu \mathrm{m}$. The flexibility of the \citet{fritzRevisitingInfraredSpectra2006} model allows for a breadth of AGN SED shapes, from totally unobscured type-I AGN to obscured type-II AGN. 

In general, we use a standard set up of \textsc{ProSpect} fitting for at most 11 parameters. For further details the reader is referred to \citet[][especially their Tab. 2]{thorneDeepExtragalacticVIsible2022}. For fitting and exploration of the posterior distribution of the parameters we use the software \textsc{Highlander}\footnote{\url{https://github.com/asgr/highlander}} that switches between 2 rounds of genetic optimisation and MCMC with the Component-wise Hit And Run Metropolis algorithm to search for the global maximum of the highly multidimensional parameter space. Final quantities are obtained from the median of the MCMC samples and uncertainties are obtained from the $16-\nth{84}$ percentiles. The right panel of Figure~\ref{fig:zphot_zspec} shows that $z_{\mathrm{phot}}$ from \textsc{ProSpect} and the \textsc{EAZY} prior agrees with the pure \textsc{EAZY} fits; though, with one fewer outlier. The optimisation and posterior sampling mean that we can put robust uncertainties on $z_{\mathrm{phot}}$ (in addition to the astrophysical quantities). 

\subsubsection{Modelling gas enshrouded LRDs}\label{sec:modellinglrds}
The physics that determines the shape of the LRD SEDs continues to be debated \citep[e.g.,][]{madauLittleRedDots2026a, asadaOriginsUVContinuum2026, rusakovLittleRedDots2026}. Whether it is appropriate to apply standard dusty torus models to LRDs is not entirely clear. It is hence insightful to explore alternative models for LRDs to see how, in particular, the AGN luminosities and stellar masses change under different prior assumptions. 

A combination of young stars and a gas enshrouded SMBH complex seems to be able to explain the `V-shape' of LRD SEDs \citep{inayoshiSpectralUniformityLittle2026}. Most of the continuum blueward of the Balmer limit is expected to be absorbed by the gaseous envelope and the emergent emission radiates similarly to a thermalised stellar photosphere that explains the rising rest-frame optical-IR rise \citep{kidoBlackHoleEnvelopes2025, asadaOriginsUVContinuum2026}. When the gaseous envelope enshrouds the SMBH almost none of the UV continuum nor any of the emission lines can be powered by the accretion disk and are, instead, powered by young, centrally concentrated stars. Some UV continuum and broad H$\alpha$ lines can originate from the accretion disk when the gas is arrayed in a stratified gaseous cocoon, in which case the emergent emission is not entirely thermalised \citep{naiduBlackHoleStar2025a,asadaOriginsUVContinuum2026}. 

High gas densities ($\mathrm{N_{H} \gtrsim 10^{9} \, cm^{-3}}$) produce non-stellar Balmer breaks, which would otherwise be ascribed to A-type stars in $\mathrm{M_{\star} \gtrsim 10^{10} \, M_{\odot}}$ galaxies; hence, these models can lower the inferred stellar mass compared to standard stellar population modelling \citep{inayoshiExtremelyDenseGas2025a}. LRDs are seemingly deficient in hot dust \citep[e.g.,][]{caseyDustLittleRed2024a,caseyUpperLimit1062025} so the gas-enshrouded models are particularly attractive in contrast to typical dusty torus models. These models can lower the inferred bolometric AGN luminosities because a significant hot dust component is not extrapolated.

The gas enshrouded model was implemented into \textsc{ProSpect} with a modified blackbody (MBB, or also referred to as a greybody) function with a Balmer break, 
\begin{equation}
    \mathrm{L_{AGN}^{MBB}}(\lambda) \propto \mathcal{H(\SI{3646}{\angstrom})} \times \lambda^{-\beta} B_{\lambda}(\mathrm{T_{eff}}),
    \label{eq:mbb}
\end{equation}
where $\mathcal{H(\SI{3646}{\angstrom})}$ is the Heaviside step function, which is unity for $\lambda \geq \SI{3646}{\angstrom}$ and zero elsewhere, and $B_{\lambda}(\mathrm{T_{eff}})$ is the pure blackbody function with an effective temperature, $\mathrm{T_{eff}}$. The blackbody function is employed because the emission is expected to be thermal after processing by the optically thick gas enshrouding the SMBH. The grey component accounts for variations in both the opacity of the medium and the source emissivity \citep{caseyFarinfraredSpectralEnergy2012}. The Balmer break component is expected because gas densities of $n_{\mathrm{H}} \approx 10^{9}-10^{11} \, \mathrm{cm^{-3}}$ permit sufficient numbers of hydrogen atoms in the $n=2$ state that can absorb almost all of the continuum blueward of the Balmer limit \citep{inayoshiExtremelyDenseGas2025a}. 

We also allow for independent dust attenuation of the SMBH complex with the \citet{charlotSimpleModelAbsorption2000b} model and re-emit the absorbed energy with \textsc{SKIRTOR} models of torus dust \citep{stalevski3DRadiativeTransfer2012,stalevskiDustCoveringFactor2016}. The free parameters for this new AGN model are $\mathrm{L_{AGN}}$, the effective temperature of the blackbody, the $\beta$ slope of the grey component and the V-band optical depth normalising coefficient of the AGN dust attenuation in the \citet{charlotSimpleModelAbsorption2000b} model. To be clear, this AGN model is included simultaneously with a stellar/host galaxy component in the fitting, in the same manner as done with \textsc{Stellar+AGN} runs throughout. 

The simple use of the MBB model smooths over much of the physics of radiative transfer through dense gaseous media. Full quasi-photospheric calculations were recently undertaken by \citet{liuSyntheticSpectralLibrary2026} who used the \textsc{tlusty} code \citep{hubenyComputerProgramCalculating1988,hubenyTLUSTYSYNSPECUserss2021} to model LRDs, just as is routinely done for grids of synthetic stellar spectral models. These parameterise the SED with mainly the effective temperature and the surface gravity, which is related to the total mass of the entire SMBH complex. We implemented the \citet{liuSyntheticSpectralLibrary2026} models into \textsc{ProSpect} in the same way as was done for the MBB model. 

We also experiment with cocoon models that differ from the quasi-photospheric, envelope models. \citet{sneppenCocoonComprehensiveExplanation2026} used the \textsc{sirocco} code \citep{matthewsSIROCCOPubliclyAvailable2025} to model the radiation field and ionizing structure of the cocoon, parametrising the spectrum with mainly the mass outflow rate, which is itself connected to the gas column density. A key distinction between this cocoon model and the quasi-photospheric, envelope one is the origin of the UV continuum. The cocoon model permits some UV escape from the accretion disk depending on the column density of the gas, whereas virtually all the UV continuum must originate from the host galaxy when using the envelope models. 

Throughout, the fiducial AGN model adopted in this paper is the standard dusty torus models of \citet{fritzRevisitingInfraredSpectra2006}. However, we further discuss different models and their application to LRDs in Section~\ref{sec:lrds}.

\subsection{Identification of AGNs from SED fitting} \label{sec:agnidentify}
We label sources as AGNs if at least $10$ per cent of the light in either the rest-frame UV ($\SI{1500}{\angstrom}$) or the integrated rest-frame $3-20 \, \mathrm{\mu m}$ originated from an AGN component, essentially encompassing both type-I and type-II AGNs. Similar thresholds at the same wavelengths have previously been used to isolate AGNs from SED fitting \citep[e.g.,][]{daleTwoparameterModelInfrared2014a,thorneDeepExtragalacticVIsible2022}. For galaxies where a significant AGN component could not be identified, either they do not host an accreting SMBH or it was not luminous enough to separate from the host galaxy emission with the data on hand. This motivated us to perform two sets of SED fitting: \textsc{Stellar$+$AGN} that includes an AGN component and \textsc{Stellar} that only accounts for starlight. The fitting range of AGN luminosity is $\mathrm{L_{AGN} \in [10^{35} - 10^{49}] \, erg \, s^{-1}}$ meaning that if $\mathrm{L_{AGN}}$ is low in comparison to the luminosity of the host galaxy, the two sets of \textsc{ProSpect} fits yield similar results. Indeed, the stellar masses and star formation rates for the sources without a significant AGN were similar between the two sets of fits. For the insignificant AGNs we simply adopted the fit parameters from the \textsc{Stellar} fits, preferring the simpler model. 

\begin{figure}
    \centering
    \includegraphics[width=\linewidth]{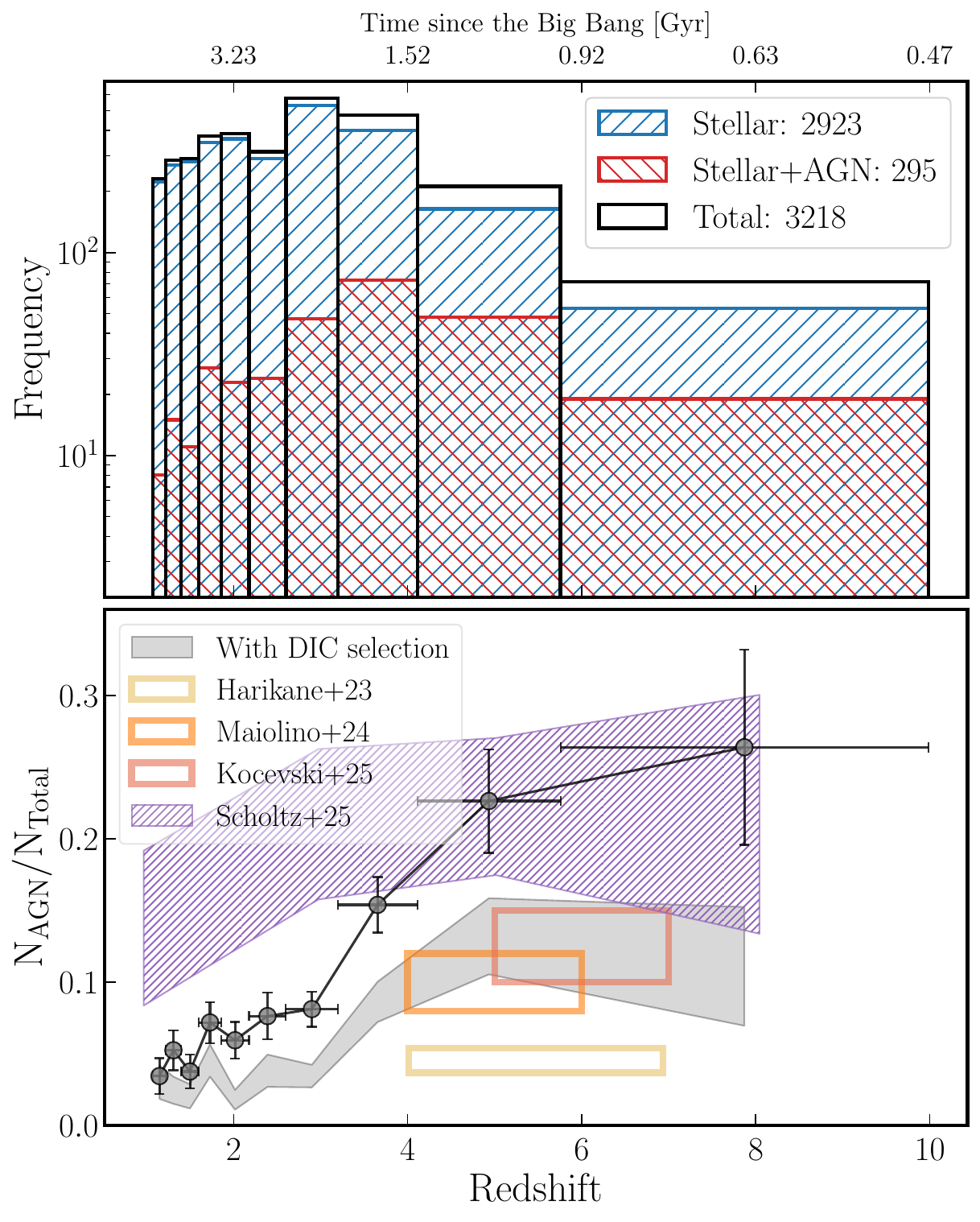}
    \caption{
    \textit{Top:} redshift histogram of our MIRI sample at $1\lesssim z \lesssim 10$. The redshift bins were chosen to span $500$~Myr of time. The black histogram shows the total sample. The blue hatched histogram shows the sources without a significant AGN component in their SEDs and the red hatched histograms show the significant AGNs. \textit{Bottom:} occupation fraction of AGNs in galaxies as a function of redshift. Points and error bars show the results and the Poisson uncertainties. The grey shaded region shows the same but with the additional DIC selection discussed in the text. The boxes show the occupation fraction of type-I AGNs reported in \citet{harikaneJWSTNIRSpecFirst2023,maiolinoJADESDiversePopulation2024,kocevskiRiseFaintRed2025a}. The purple hatched region shows the results of type-II AGNs from \citet{scholtzJADESLargePopulation2025}.
    }
    \label{fig:redshift_hist}
\end{figure}
The top panel of Figure~\ref{fig:redshift_hist} shows the redshift distribution of our final $1 \lesssim z_{\mathrm{phot}} \lesssim 10$ sample. In total, there are $3218$ galaxies, $295$ of which we identify as AGNs. The sample was binned in redshift such that each bin covered $500$~Myr of time. The bottom panel of Figure~\ref{fig:redshift_hist} shows the occupation fraction of AGNs for all objects in the sample as a function of redshift from our methods of identification discussed above. We find that at $z\lesssim3$ the occupation fraction of AGNs is stable at $< 0.1$, before rapidly increasing up to $z\approx 8$ where it is $\approx 0.25$. 

We compared the occupation fraction of AGNs in galaxies to results from NIRSpec spectroscopically confirmed type-I H$\alpha$ broad-line AGNs \citep{harikaneJWSTNIRSpecFirst2023,maiolinoJADESDiversePopulation2024} and a sample of photometrically identified LRDs \citep{kocevskiRiseFaintRed2025a}. It can be seen that we estimate $\approx 2$ times more AGNs. 

Our measurements for the occupation fraction of AGNs are similar to those reported by \citet{scholtzJADESLargePopulation2025} who search for type-II AGNs through high ionisation lines from NIRSpec observations in the JADES survey. There is some tension however between our AGN occupation fractions and those from \citet{scholtzJADESLargePopulation2025} at $z \lesssim 4$ that likely reflects differences in the AGN selection methods. \citet{scholtzJADESLargePopulation2025} select $\approx 40$ AGNs from a parent sample of $\approx 200$ JADES spectra. Their AGN occupation fraction therefore reflects the number of AGNs in the parent sample and not necessarily in the average population of galaxies.

\begin{figure}
    \centering
    \includegraphics[width=\linewidth]{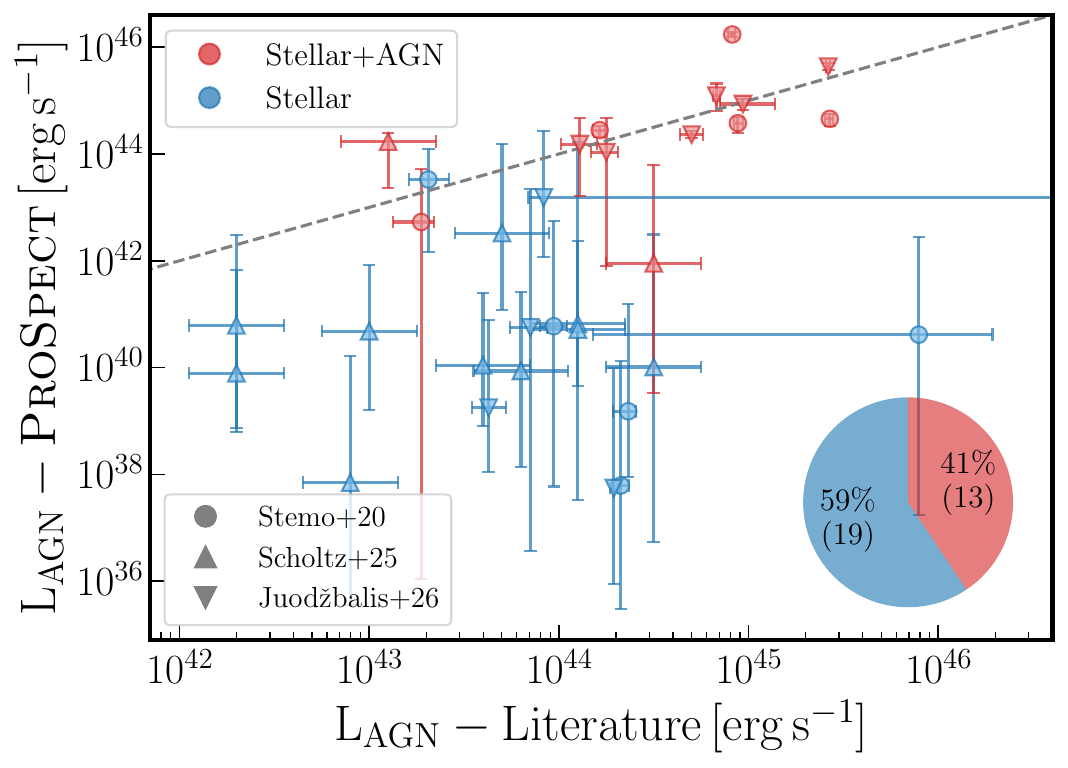}
    \caption{
    Bolometric luminosities of AGN inferred from \textsc{ProSpect} against those reported in the literature at $1 \lesssim z \lesssim 10$ (10 X-ray AGNs from \citealt{stemoCatalogAGNHost2020}, 12 type-II NIRSpec AGNs from \citealt{scholtzJADESLargePopulation2025}, 10 type-1 AGNs from \citealt{juodzbalisJADESComprehensiveCensus2026}). There are 32 sources from a coordinate match with our photometric catalogue. The red points and error bars show the galaxies we estimate as significant AGNs and the blue points show the galaxies where the AGN contributed $<10$ per cent to the SED as per our significant AGN selection. The dashed line shows the one-to-one relation. The pie chart shows the proportion of \textsc{Stellar} and \textsc{Stellar$+$AGN} sources from the matched sample. $13$ out of $32$ sources are robust AGNs as identified from both our SED fitting and independently from the literature. 
    }
    \label{fig:lagn_comparison}
\end{figure}

\citet{scholtzJADESLargePopulation2025} also probe faint ($\mathrm{L_{AGN} \lesssim 10^{44} \, erg \, s^{-1}}$) AGNs that may have been missed in our sample due to the aforementioned challenge of separating AGN and host galaxy contributions to the SED. To further expand on this, we compare in Figure~\ref{fig:lagn_comparison} $\mathrm{L_{AGN}}$ obtained from our SED fitting to that from \textit{Chandra} $2-10 \, \mathrm{keV}$ X-ray observations at $z<2.5$ \citep{stemoCatalogAGNHost2020}, type-I broad H$\alpha$ components from NIRSpec at $1.5 < z< 9$ \citep{juodzbalisJADESComprehensiveCensus2026} and the same NIRSpec type-II AGNs from \citet{scholtzJADESLargePopulation2025}. A coordinate match between these literature catalogues and our own yielded 32 objects. 

It is clear that we recover similar values of $\mathrm{L_{AGN}}$ to those in the literature for $\mathrm{L_{AGN}} \gtrsim 10^{44} \, \mathrm{erg \, s^{-1}}$. Below this critical value, $\mathrm{L_{AGN} \approx 10^{44} \, erg \, s^{-1}}$, we struggle to detect a significant AGN component from SED fitting with the broadband data on hand, owing to the difficulty in separating the AGN emission from the host component. Per unit of stellar mass, or otherwise referred to as the specific AGN luminosity, this critical value corresponds to $\mathrm{L_{AGN}/M_{\star} \approx 10^{34} \, erg \, s^{-1} \, M_{\odot}}$. 

To be clear, in Figure~\ref{fig:lagn_comparison} we show $\mathrm{L_{AGN}}$ obtained from the \textsc{Stellar$+$AGN} fit throughout but separate sources as significant AGN or not. It should be noted that both the X-ray and spectroscopic results rely on possibly uncertain bolometric corrections \citep[e.g.,][]{netzerBolometricCorrectionFactors2019a,greeneWhatYouSee2026}. Additionally, the spectroscopic results rely on successful separation of stellar and AGN powered narrow-line emission \citep[e.g.,][]{feltreNuclearActivityStar2016a}, which could mean that the inferred $\mathrm{L_{AGN}}$ values are upper limits \citep{scholtzJADESDataRelease2026}. 

Of the 32 matches, 13 ($\approx 41$ per cent) sources have multiple lines of evidence of hosting an AGN, from broad-line/X-ray selections and our own SED fitting. 10 out of the 32 sources were independently identified as broad-line sources with FWHM$>900 \, \mathrm{km\,s^{-1}}$ in H$\alpha$ or H$\beta$ from R1000 NIRSpec observations. Three of these sources were classed as LRDs and two as Little Blue Dots (LBDs) because they are compact in the F444W band and have UV/optical continuum slopes consistent with the definitions in \citet{kocevskiRiseFaintRed2025a,taylorBroadlineAGNs352025,brazziniLittleBlueRed2026a}. We photometrically identified 2/3 LRDs, 2/2 LBDs and 3/5 of the remaining broad-line sources as AGNs. 

This comparison between our photometric sample and those from the literature are subject to selection effects. The AGNs identified by X-rays from \citet{stemoCatalogAGNHost2020} are limited to significant detections only at $z<2.5$. The AGNs identified by NIRSpec from \citet{juodzbalisJADESComprehensiveCensus2026} and \citet{scholtzJADESDataRelease2026} are representative of the targets selected for spectroscopic follow-up and not necessarily the average population of AGNs. Our own selection relies on the fractional contribution of AGN to the rest-frame UV and integrated $3-20 \, \mathrm{\mu m}$. It also depends on the F560W and F770W bands for the source detection, which probe $\lambda_{\mathrm{rest}} \approx 2.8/3.8 \, \mu \mathrm{m}$ at $z=1$ and $\lambda_{\mathrm{rest}} \approx 0.5/0.7 \, \mu \mathrm{m}$ at $z=10$. We advocate for multiband spectro-photometric searches to mitigate selection effects and improve AGN constraints from spectral fitting. 

Including the AGN component in \textsc{ProSpect} introduces 3 extra free parameters in addition to the 8 required to model the redshift, stars and dust. A common strategy for selecting AGN is to apply hypothesis testing to assess whether the extra degrees of freedom of an AGN included SED fit is warranted to characterise given the data \citep[e.g.,][]{carranza-escuderoLonelyLittleRed2025}. Hence, we experimented with adding a statistical requirement to our AGN selection. We used the Deviance Information Criterion (DIC), a generalisation of the Akaike Information Criterion for Bayesian model selection, that leverages information about the entire posterior distribution \citep{gelman2004bayesian}. Specifically, between \textsc{Stellar} and \textsc{Stellar+AGN} we preferred the model with lower DIC (i.e., DIC$_{\mathrm{Stellar+AGN}}$ - DIC$_{\mathrm{Stellar}}$ $< 0$). With this requirement, in addition to the SED fractional contribution requirements, we found that the occupation fraction of AGNs that we detected decreased by about half, coming closer into agreement with the type-I results from \citet{harikaneJWSTNIRSpecFirst2023,maiolinoJADESDiversePopulation2024,kocevskiRiseFaintRed2025a}. The comparison of $\mathrm{L_{AGN}}$ against the broad H$\alpha$ results also showed that even when values $\mathrm{ L_{AGN} \gtrsim 10^{44} \, erg \, s^{-1}}$ agreed the DIC selection preferred the \textsc{Stellar} fits for some sources. The challenging aspect of selecting AGNs in this manner is to essentially balance completeness against purity in our sample. We therefore adopted the looser criteria without the DIC requirement to avoid missing significant AGNs with $\mathrm{ L_{AGN} \gtrsim 10^{44} \, erg \, s^{-1}}$. 

As a test, we refit the 13 robust AGNs shown in Figure~\ref{fig:lagn_comparison} with only their HST and NIRCam measurements. In this case, only 3 of the 13 sources were photometrically identified as significant AGNs, demonstrating the necessity of MIRI. The stellar masses were unchanged and the SFRs were $\approx 0.1$~dex lower when including MIRI, which is consistent with similar findings by \citet{liEPOCHSPaperVIII2024}. 

\section{Demographics of MIRI selected AGNs} \label{sec:agndemographics}
\begin{figure*}
    \centering
    \includegraphics[width=\textwidth]{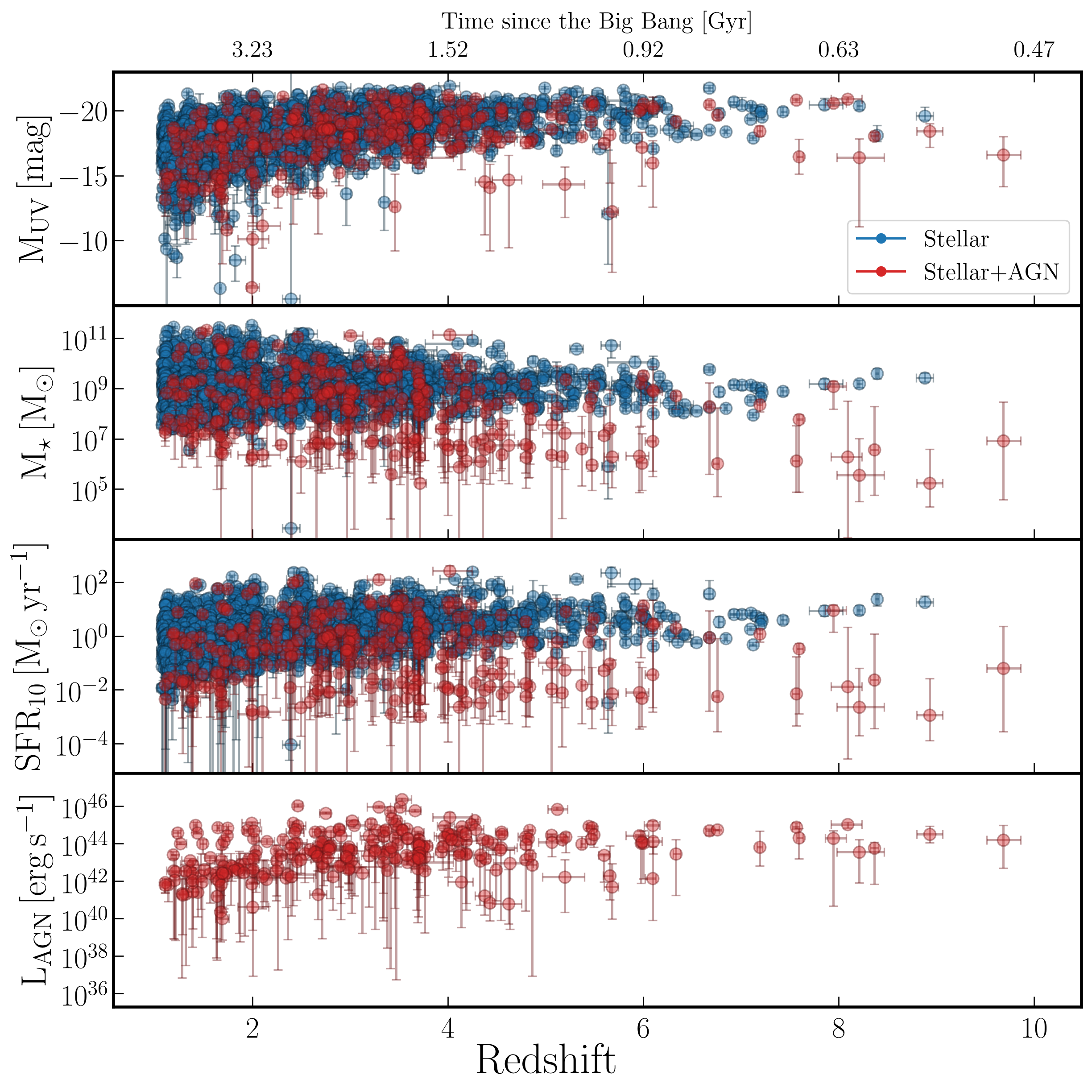}
    \caption{$\mathrm{M_{UV}}$ (\textit{top row}), $\mathrm{M_{\star}}$ (\textit{second row}), $\mathrm{SFR_{10}}$ (\textit{third row}) and $\mathrm{L_{AGN}}$ (\textit{bottom row}) against redshift. $\mathrm{M_{UV}}$ is measured from a tophat filter of width $\SI{100}{\angstrom}$ and centred on $\SI{1500}{\angstrom}$. The SFR was inferred by averaging the SFH over the last $10$~Myr. In all panels, we divide galaxies between \textsc{Stellar+AGN} and \textsc{Stellar} as indicated in the figure legend.}
    \label{fig:astro_redshift}
\end{figure*}
Figure~\ref{fig:astro_redshift} shows the key astrophysical quantities from our sample against redshift. We removed 13 sources that had either $\mathrm{M_{\star} \geq 10^{12} \, M_{\odot}}$ or $\mathrm{M_{UV} \leq -22 \, \mathrm{mag}}$. Visual inspection showed that these sources suffered from flux contamination from neighbouring sources, erroneously boosting the observed fluxes and resulting stellar mass/$\mathrm{M_{UV}}$. In any case real sources that satisfied $\mathrm{M_{\star} \geq 10^{12} \, M_{\odot}}$ or $\mathrm{M_{UV} \leq -22 \, \mathrm{mag}}$ are unlikely to have been observed in our volume-limited sample over $\approx 35 \, \mathrm{arcmin}^{2}$. 

The SFR was inferred by averaging the fitted SFH over the last $10$~Myr. We compared our $\mathrm{SFR_{10}}$ values with a cross match of 36 H$\alpha$ measurements from NIRCam/grism F444W spectra from the FRESCO survey at $4.9 < z < 6.7$ \citep{oeschJWSTFRESCOSurvey2023,covelo-pazHaViewGalaxy2025}. The H$\alpha$ fluxes were first corrected for dust attenuation using an unchanging attenuation value of $A_{\mathrm{H}\alpha} = 0.47$, which is the median of the sample reported in \citet{covelo-pazHaViewGalaxy2025}, and then converted to SFRs using the conversion factor ($\mathrm{SFR_{H\alpha} = 10^{-41.36}L_{H\alpha}[erg \, s^{-1}]}$) appropriate for a \citet{chabrierGalacticStellarSubstellar2003} IMF. This showed that the H$\alpha$ SFRs were $\approx 0.17$~dex higher than our own $\mathrm{SFR_{10}}$ values; though, this will depend on the dust correction value for the H$\alpha$ fluxes.

Across redshift, $\mathrm{SFR_{10}}$ is generally bound between values of $\mathrm{10^{-3} - 10^{3} \, M_{\odot} \, yr^{-1}}$. However, quenched galaxies are defined with the following criterion,
\begin{equation}
    \mathrm{sSFR} < 0.2/t_{U}(z),
    \label{eq:quenching}
\end{equation} where $\mathrm{sSFR \, [yr^{-1}] = SFR_{10}/M_{\star}}$ and $t_{U}(z)$ is the age of the Universe at the epoch of observation \citep{pacificiEVOLUTIONSTARFORMATION2016,russellCosmicStillnessHigh2025a}, and this threshold is used to establish an upper limit on the SFRs. 

It is clear from Figure~\ref{fig:astro_redshift} that many of the galaxies with significant AGN populate the lower locus of both stellar mass, $\mathrm{M_{\odot} \lesssim 10^{9} \, M_{\odot}}$ and $ \mathrm{SFR_{10}} \lesssim 10^{-2} \, \mathrm{M_{\odot} \, yr^{-1}}$. This is because the SED is dominated by the AGN component and there is little allowance for a significant contribution of starlight. 

\subsection{Star forming main sequence} \label{sec:SFMS}
\begin{figure*}
    \centering
    \includegraphics[width=\textwidth]{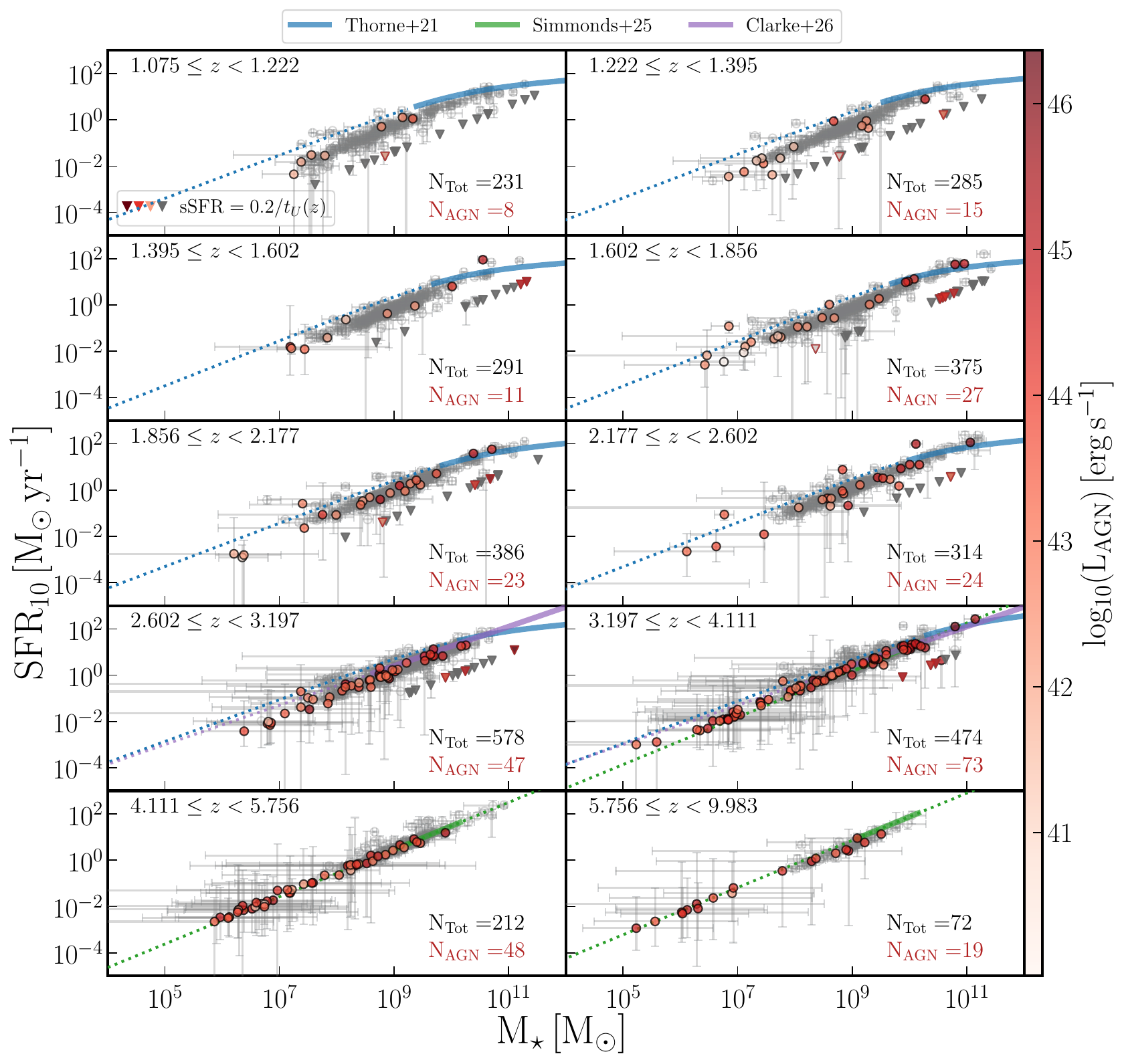}
    \caption{In every panel, we show the $\mathrm{SFR_{10} - M_{\star}}$ in each redshift bin as indicated. Error bars on the points show the 1$\sigma$ uncertainties from the posterior distributions. Filled points show significant AGN with the colour associated with the inferred $\mathrm{L_{AGN}}$ as indicated with the colour bar. Downward facing triangles show the critical values of $\mathrm{SFR_{10}}$ as calculated from our definition of quenching in Equation~\ref{eq:quenching}. Dotted and dashed lines show fits to the relation from \citet{thorneDeepExtragalacticVIsible2021,simmondsBurstingSeamsStarforming2025,clarkeStarformingMainSequence2026a} as indicated in the figure legend. The dotted line for these literature fits show the extrapolations, beyond the reported stellar mass completeness limits.}
    \label{fig:SFMS}
\end{figure*}

To further demonstrate this, we show in Figure~\ref{fig:SFMS} the $\mathrm{SFR_{10} - M_{\star}}$ relation in each redshift bin for our sample and colour the points by $\mathrm{L_{AGN}}$. Especially at $z\gtrsim 4$, the galaxies with the lowest stellar masses, $\mathrm{M_{\star} \lesssim 10^{9} \, M_{\odot}}$, have dominant AGN contributions as indicated by their $\mathrm{L_{AGN}}$. 

At all redshifts, we observe correlations between $\mathrm{SFR_{10}-M_{\star}}$ that appears to persist down to $\mathrm{\approx 10^{5} - 10^{7} \, M_{\star}}$ depending on the redshift, finding agreement with previous (extrapolated) fits of the relations from \citet{thorneDeepExtragalacticVIsible2021,simmondsBurstingSeamsStarforming2025,clarkeStarformingMainSequence2026a}; and hence encouraging us on the performance of our SED fitting and SFH constraints, in the context of other, largely independent methods. 

The incidence of AGNs at $\mathrm{M_{\star} \lesssim  10^{9} \, M_{\odot}}$ may be symptomatic of the inability to recover a significant host component, the complement of the issue highlighted in Figure~\ref{fig:lagn_comparison}, and known as AGN outshining \citep[e.g.,][]{marshallObservingHostGalaxies2021,juodzbalisEPOCHSVIIDiscovery2023}. 

The ratio of the number of galaxies hosting AGN to the total in this stellar mass range can be as high as $\approx 1$, especially in the highest redshift bins. If the host component is truly subdominant then this indicates that a significant proportion of the luminosity was powered by AGN accretion in comparison to star formation. 

\begin{figure*}
    \centering
    \includegraphics[width=\textwidth]{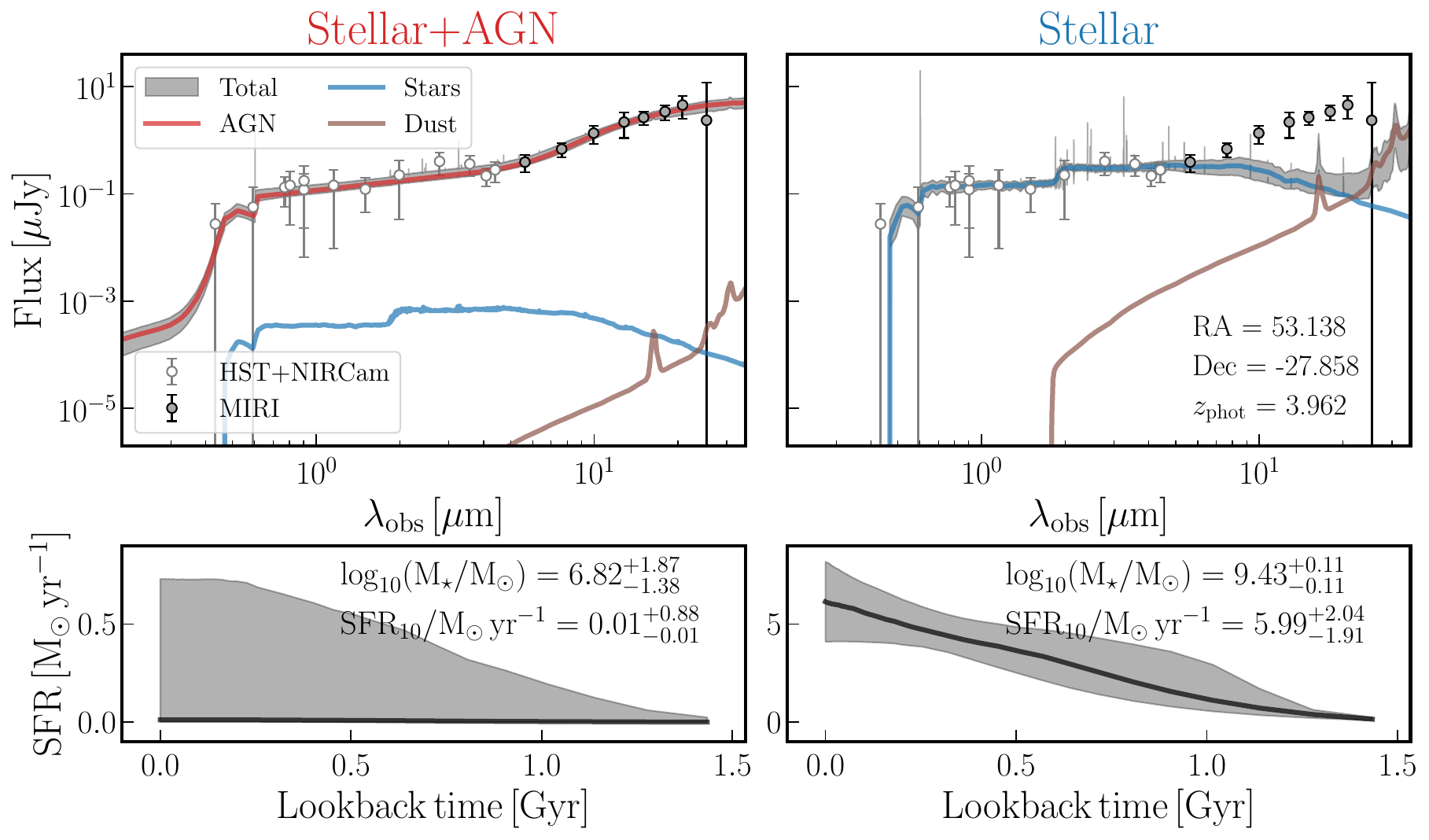}
    \caption{Example SED fits of a $\mathrm{M_{\star} < 10^{8} \, M_{\odot}}$ galaxy better fit by the \textsc{Stellar+AGN} model. \textit{Left:} \textsc{Stellar+AGN} SED fit. The points with error bars show the measured and 1$\sigma$ uncertainties of the photometry. The black filled region shows the total SED fit and $16$-\nth{84} quantiles. The red line shows median AGN component, the blue line shows the median stellar component and the brown line shows the median dust component. \textit{Right:} same as the left panel but for the \textsc{Stellar} SED fit. The sky coordinates and photometric redshift are indicated in the panel. \textit{Bottom:} star formation histories between SED fits. We highlight the median and $16$-\nth{84} uncertainties of the SFR$_{10}$ and the stellar mass. $\mathrm{Lookback \, time=0}$ corresponds to the epoch of observation.}
    \label{fig:galaxy_sed}
\end{figure*}
For these low stellar mass galaxies that are mostly dominated by an AGN component, there is little room in the SED fit for a significant host component. This can be seen in Figure~\ref{fig:galaxy_sed} where we compare the SED fits between \textsc{Stellar+AGN} and \textsc{Stellar} for a galaxy selected as dominated by an AGN and with $\mathrm{M_{\star} < 10^{8} \, M_{\odot}}$. The posterior samples of the SFH for the \textsc{Stellar+AGN} fit are largely unconstrained and the median SFH is virtually constant over the formation history. This results in the log-linear correlation between the median $\mathrm{SFR_{10} - M_{\star}}$ values and large uncertainties. It is clear that the AGN component is necessary to account for the rest-frame optical-IR flux because a similar hot, $\mathrm{T\sim 1000 K}$, dust component from an AGN is not accounted for in the \textsc{Stellar} only model. 

A subdominant host component is indicative of rapid SMBH assembly, outpacing the growth of the stellar mass in $\mathrm{M_{\star} \lesssim 10^{9} \, M_{\odot}}$ galaxies. There is evidence of dwarf galaxies hosting an AGN at $z \lesssim 3$ \citep{mezcuaOvermassiveBlackHoles2023,mezcuaOvermassiveBlackHoles2024}. Similarly, over-massive SMBHs, $\gtrsim 1$~dex above local $\mathrm{M_{\star} - M_{\bullet}}$ relations \citep{reinesRelationsCentralBlack2015,greeneIntermediateMassBlackHoles2020a}, appear to be ubiquitous at $z\gtrsim 4$ \citep{kokorevUNCOVERNIRSpecIdentification2023,furtakHighBlackholetohostMass2024,maiolinoJADESDiversePopulation2024,juodzbalisJADESComprehensiveCensus2026}. Hence, these objects are found across redshift despite the decreasing availability of the gas supply and changing conditions of both the CGM and ISM with decreasing redshift \citep{tumlinsonCircumgalacticMedium2017}. Interestingly, while some $z\gtrsim 4$ SMBHs are over-massive compared to their host galaxy's stellar mass, there is evidence of agreement between the high redshift and local $\mathrm{M_{dynamical} - M_{\bullet}}$ relations \citep{kormendyCoevolutionNotSupermassive2013,izumiSubaruHighzExploration2021,maiolinoJADESDiversePopulation2024} indicating that the gas is preferentially consumed through SMBH accretion than it is from star formation. 

The AGNs themselves must have provided a source of feedback against star formation, for example through driving both neutral and ionised gas outflows \citep{daviesIonizedOutflowsLocal2020,daviesJWSTRevealsWidespread2024}, which there is evidence of in dwarf galaxies at $z \approx 0.5-0.9$ \citep{vietriTypeIIAGNhost2022,siudekEnvironmentAGNDwarf2023}. From our definition of quenching in Equation~\ref{eq:quenching}, it is clear in Figure~\ref{fig:SFMS} that the proportion of quenched galaxies increases with decreasing redshift. We observe in Figure~\ref{fig:SFMS} that as the stellar mass builds up at later times AGNs are more likely to occupy the upper-right locus of the SFMS and inhibit star formation \citep[e.g.,][]{katsianisEvolvingMassdependentSsSFRM2019,daviesGalaxyMassAssembly2019,daviesDeepExtragalacticVIsible2022,daviesDeepExtragalacticVIsible2025b}. 

Indeed, AGN feedback is a key physical mechanism invoked to explain the occurrence of quenched $\mathrm{M_{\star} \gtrsim 10^{10} \, M_{\odot}}$ galaxies at $z \approx 3$ \citep[e.g.,][]{lagosDiverseStarFormation2025,chandro-gomezUnveilingPopulationMassive2026}. AGN feedback is deleterious to itself as the accretion of matter onto the SMBH is likewise inhibited as is the formation of stars; the cosmic black hole accretion rate history declines at $z\lesssim 2$. Consistent with this, we see evidence of cosmic downsizing of AGNs as their luminosities decrease by $\approx 2$~dex in the time between the highest redshift bin to the lowest. This foreshadows a population-averaged transition from the radiatively efficient quasar mode of accretion to the more sedate radio mode \citep{fanidakisEvolutionActiveGalactic2012}. 

Especially at $z\gtrsim 2$, the ratio of the number of galaxies hosting a significant AGN to the total is $\lesssim 0.1$ in the stellar mass range $\mathrm{M_{\star} \geq 10^{9}\,M_{\odot}}$. As mentioned previously, this could mean that the host galaxy outshines the AGN and we could not significantly detect it with the data on hand. Alternatively, for $\gtrsim 90$~per cent of these $\mathrm{M_{\star}} \geq 10^{9}\,\mathrm{M_{\odot}}$ galaxies, the SMBH may simply not have been observed in a luminous phase, indicative of lower SMBH accretion rates in these massive galaxies at the instant of observation. 

The inhibition of gas conversion due to AGN feedback is the cumulative effect of SMBH growth and the key quantity is therefore the SMBH mass, as opposed to the instantaneous $\mathrm{L_{AGN}}$ \citep{bluckQuenchingGalaxiesBulges2022,piotrowskaQuenchingStarFormation2022}. As the SMBH mass broadly scales with the stellar mass \citep[e.g.,][]{kormendyCoevolutionNotSupermassive2013}, we expect AGN feedback to be more prominent in comparatively more massive galaxies. 

The typical duration of the luminous phase of quasars is $t_{\mathrm{QSO}} \lesssim 10^{7} \, \mathrm{yr}$ \citep[e.g.,][]{eilersDetectingCharacterizingYoung2020,khrykinFirstMeasurementQuasar2021} and our redshift bins cover $\approx 500$~Myr. Naively comparing these two timescales means that the typical duty cycle would be $10/500 \lesssim 0.1$, which is comparable to the AGN occupation fraction in these $\mathrm{M_{\star} \geq 10^{9}\,M_{\odot}}$ galaxies.

\subsection{Growth of SMBHs} \label{sec:growthsmbhs}
As the stellar and SMBH mass grow fundamentally from the same gas reservoir in galaxies, it is insightful to compute the dimensionless baryon conversion ratio. The black hole accretion rate (BHAR) is related to the AGN luminosity as, 
\begin{equation}
    \label{eq:emc2}
    \mathrm{BHAR} = \frac{\mathrm{L_{AGN}}}{\epsilon c^{2}},
\end{equation}
where $\epsilon$ is the radiative efficiency and $c$ is the speed of light. The radiative efficiency is $\epsilon \approx 0.06  - 0.4$, increasing with the spin of the SMBH for standard thin-disk accretion \citep{shakuraBlackHolesBinary1973,thorneDiskAccretionBlackHole1974a}, and the canonical value frequently quoted is $\epsilon \approx 0.1$ \citep[e.g.,][]{soltanMassesQuasars1982,davisRadiativeEfficiencyAccretion2011}. We uniformly sample values of $\epsilon$ between $0.06-0.4$ to account for this uncertainty in the computation of $\mathrm{BHAR}$.  

\begin{figure*}
    \centering
    \includegraphics[width=\linewidth]{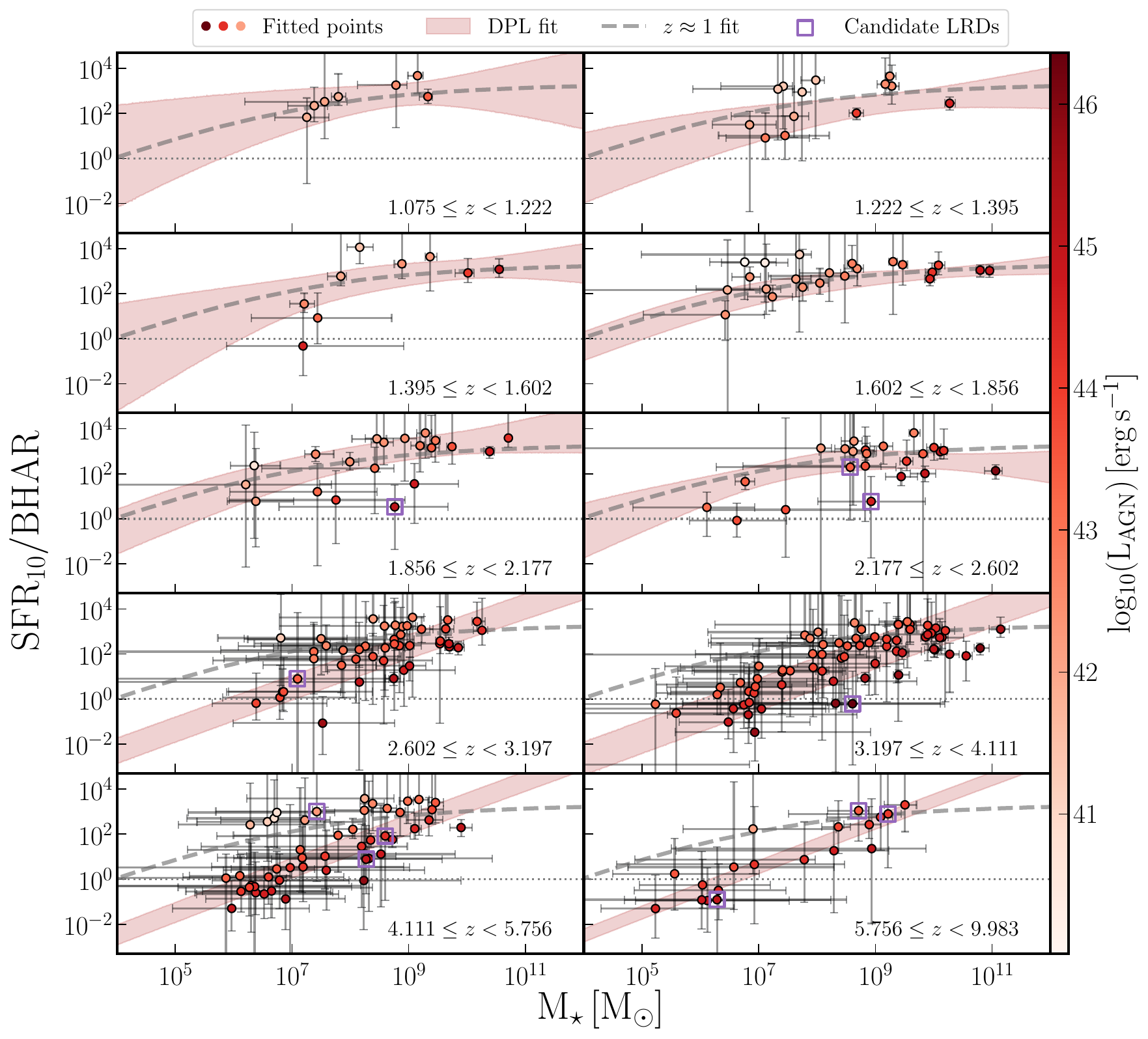}
    \caption{In every panel we show the $\mathrm{SFR_{10}/BHAR - M_{\star}}$ in each redshift bin as indicated. Error bars on the points show the 1$\sigma$ uncertainties from the posterior distributions. The colour of each point is associated with the inferred $\mathrm{L_{AGN}}$ as indicated with the colour bar. Filled lines show the 16-\nth{84} contours (plus the fitted intrinsic scatter) of the double power-law fits to the relations from Equation~\ref{eq:dbl}. The dashed line in every panel is the median fit at $z\approx 1$. Points with purple boxes around them mark LRD candidates, discussed in Section~\ref{sec:lrds}. The dotted line in every panel highlights $\mathrm{SFR_{10} = BHAR}$.}
    \label{fig:bhar_sfr}
\end{figure*}
Figure~\ref{fig:bhar_sfr} shows the dimensionless baryon conversion ratio between stars and SMBHs as a function of stellar mass, with the BHAR values calculated as per Equation~\ref{eq:emc2} and the fitted $\mathrm{L_{AGN}}$ values. Ratios below unity indicate that the growth of the SMBH outpaces the assembly of stellar mass, and vice versa for ratios above one. 

To quantify this, we fitted double power-law functions, of the form, 
\begin{equation}
    \label{eq:dbl}
    \mathrm{\frac{SFR_{10}}{BHAR}(\mathcal{M})} = \frac{S_{0}}{
    \left( \frac{10^{\mathcal{M}}}{10^{\mathcal{M}_{0}}} \right)^{-\alpha} + 
    \left( \frac{10^{\mathcal{M}}}{10^{\mathcal{M}_{0}}} \right)^{-\beta}
    },
\end{equation}
where $\mathcal{M} = \log_{10}(\mathrm{M_{\star}})$, $\alpha$ is the low mass slope, $\beta$ is the high mass slope, $\mathcal{M}_{0}$ is the break mass and $S_{0}$ is the asymptotic value of the function with increasing $\mathcal{M}$. We also include an intrinsic scatter parameter in $\mathrm{\frac{SFR_{10}}{BHAR}(\mathcal{M})}$. A fully Bayesian approach was followed that incorporated uncertainties on both the stellar mass and $\mathrm{\frac{SFR_{10}}{BHAR}}$ variables, which were themselves obtained from the \textsc{ProSpect} posterior distributions. 

At $z \approx 1-2$, most galaxies populate $\mathrm{\frac{SFR_{10}}{BHAR} \gtrsim 1}$ that is consistent with the peak of the cosmic star formation rate density. While this epoch is also the peak of the cosmic black hole accretion rate density, SMBHs are $\approx 1000$ times less space dense than stars per unit comoving volume. 
\begin{figure}
    \centering
    \includegraphics[width=\linewidth]{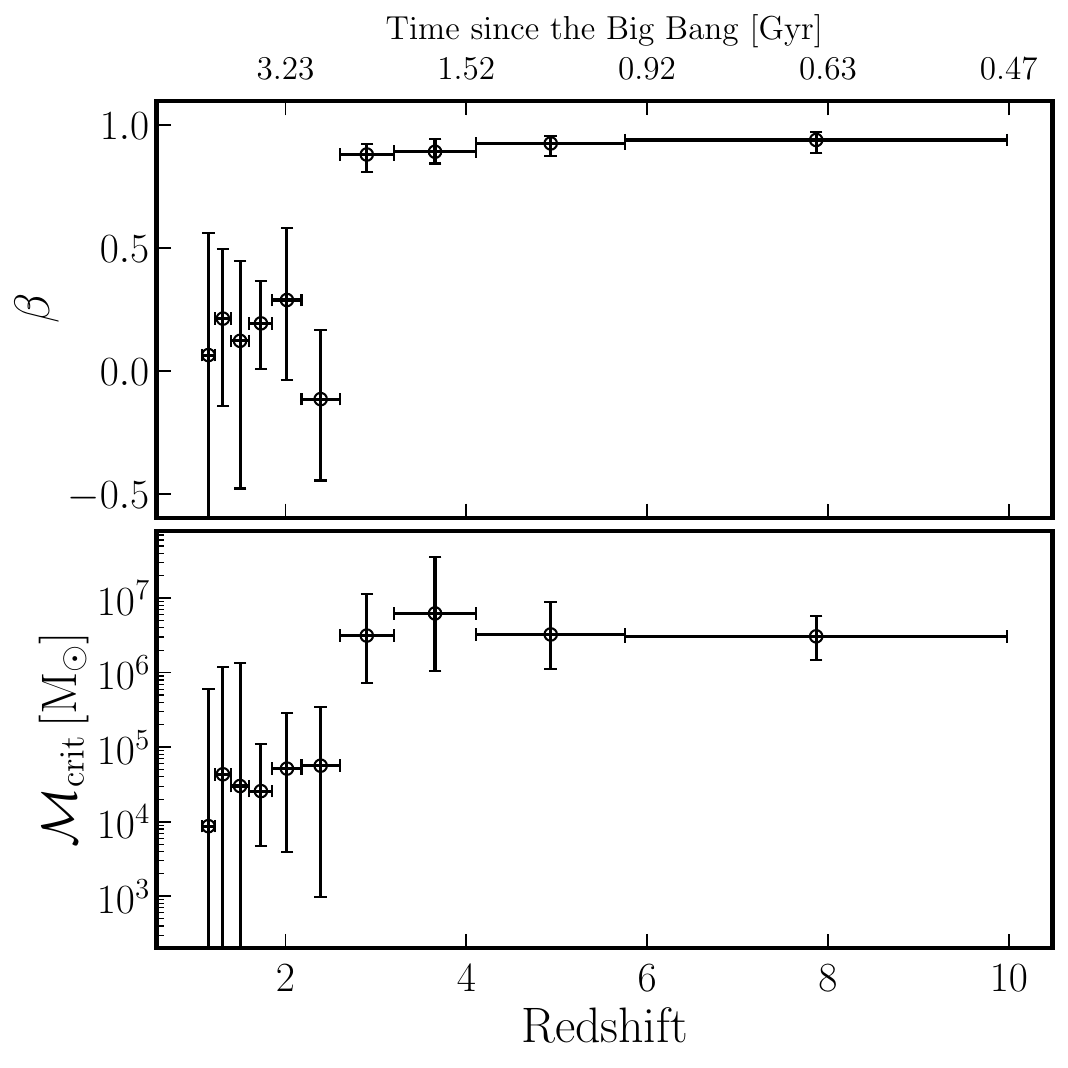}
    \caption{\textit{Top:} redshift evolution and 1$\sigma$ uncertainties of $\beta$, the high stellar mass slope. $\beta \approx 1$ at $z>3$ indicates that the fit to $\mathrm{\frac{SFR_{10}}{BHAR} - M_{\star}}$ is largely log-linear, as seen in Figure~\ref{fig:bhar_sfr}. \textit{Bottom:} redshift evolution of the critical mass, $\mathcal{M}_{\mathrm{crit}}$, at which $\mathrm{SFR_{10} = BHAR}$. These quantities are from Equation~\ref{eq:dbl}.}
    \label{fig:sfr_bhar_evol}
\end{figure}

Figure~\ref{fig:sfr_bhar_evol} shows the redshift evolution of the high mass slope, $\beta$, and the critical mass, $\mathcal{M}_{\mathrm{crit}}$, at which $\mathrm{SFR_{10} = BHAR}$. The high mass $\beta$ slope tends to flatten at $z \approx 1-2$. Bending/flattening out is also observed in the SFMS in similar stellar mass regimes because AGN feedback drives down the SFRs of galaxies \citep{popessoMainSequenceStarforming2019,thorneDeepExtragalacticVIsible2021}. At $z\gtrsim 3$, the fits are basically log-linear with no bending. 

We find a critical stellar mass of $\mathcal{M}_{\mathrm{crit}} \approx 10^{7} \, \mathrm{M_{\odot}}$ at $z \gtrsim 3$, indicating that AGN activity is dominant in galaxies with $\mathrm{M_{\star} \lesssim \mathcal{M}_{crit}}$. At $z<3$, this critical mass generally shifts to lower values, beyond what is probed by the data. The redshift evolution of all the fitted parameters are reported in Table~\ref{tab:dpl_params}.
\begin{table*}
\centering
\begin{tabular}{lcccccc}
$\langle z \rangle$ & $\mathrm{ \mathcal{M}_{0} \, [\log_{10}(M_{\odot})] }$ & $\mathrm{ \alpha }$ & $\mathrm{ \beta }$ & $\mathrm{ S_{0}}$  & $\mathrm{ \mathcal{M}_{crit} \, [\log_{10}(M_{\odot})] }$ & $\mathrm{ \sigma^{2}_{intrinsic} }$
\\
\hline
1.15 & $7.48^{+2.37}_{-2.42}$ & $0.89^{+0.66}_{-0.62}$ & $0.06^{+0.50}_{-0.80}$ & $2.83^{+0.89}_{-1.44}$ & $3.94^{+1.85}_{-8.07}$ & $0.06^{+0.27}_{-0.05}$ \\
1.31 & $6.99^{+2.48}_{-2.34}$ & $0.89^{+0.44}_{-0.35}$ & $0.21^{+0.29}_{-0.36}$ & $2.14^{+0.93}_{-1.22}$ & $4.64^{+1.44}_{-2.52}$ & $0.07^{+0.29}_{-0.06}$ \\
1.50 & $7.80^{+2.72}_{-2.44}$ & $0.94^{+0.81}_{-0.53}$ & $0.12^{+0.33}_{-0.60}$ & $2.83^{+0.87}_{-1.35}$ & $4.48^{+1.65}_{-4.05}$ & $0.05^{+0.20}_{-0.04}$ \\
1.73 & $6.53^{+1.10}_{-1.50}$ & $1.00^{+0.04}_{-0.04}$ & $0.19^{+0.17}_{-0.19}$ & $2.22^{+0.71}_{-1.02}$ & $4.41^{+0.63}_{-0.73}$ & $0.02^{+0.05}_{-0.01}$ \\
2.02 & $6.87^{+1.54}_{-1.93}$ & $1.09^{+0.20}_{-0.13}$ & $0.29^{+0.29}_{-0.32}$ & $2.51^{+0.96}_{-1.20}$ & $4.71^{+0.74}_{-1.12}$ & $0.10^{+0.32}_{-0.08}$ \\
2.39 & $7.87^{+0.95}_{-2.03}$ & $0.98^{+0.35}_{-0.35}$ & $-0.11^{+0.28}_{-0.33}$ & $2.78^{+0.44}_{-0.83}$ & $4.75^{+0.79}_{-1.77}$ & $0.08^{+0.16}_{-0.07}$ \\
2.90 & $7.26^{+1.75}_{-1.90}$ & $0.98^{+0.03}_{-0.03}$ & $0.88^{+0.04}_{-0.07}$ & $1.02^{+1.61}_{-1.68}$ & $6.50^{+0.56}_{-0.64}$ & $0.26^{+0.19}_{-0.13}$ \\
3.65 & $7.36^{+2.07}_{-3.63}$ & $0.99^{+0.03}_{-0.02}$ & $0.89^{+0.05}_{-0.05}$ & $0.87^{+1.95}_{-3.40}$ & $6.79^{+0.75}_{-0.77}$ & $0.52^{+0.21}_{-0.18}$ \\
4.93 & $8.49^{+1.89}_{-1.88}$ & $1.01^{+0.02}_{-0.02}$ & $0.93^{+0.03}_{-0.05}$ & $2.24^{+1.83}_{-1.83}$ & $6.51^{+0.44}_{-0.46}$ & $0.18^{+0.13}_{-0.08}$ \\
7.87 & $7.01^{+2.29}_{-1.80}$ & $1.05^{+0.04}_{-0.03}$ & $0.94^{+0.03}_{-0.05}$ & $0.83^{+2.21}_{-1.80}$ & $6.49^{+0.27}_{-0.32}$ & $0.03^{+0.07}_{-0.02}$ \\
\hline
\end{tabular}
\caption{Best-fit double power law parameters, as per Equation~\ref{eq:dbl}, at each redshift. Values are posterior medians with 16-\nth{84} percentile uncertainties. $\mathrm{ \mathcal{M}_{crit} }$ is the stellar mass at which $\mathrm{SFR_{10} = BHAR}$.} 
\label{tab:dpl_params}
\end{table*}

\subsection{Black hole masses} \label{sec:bhmasses}
The occurrence of over-massive SMBHs would have required considerable growth of the progenitor black holes within the first $\approx 1$~Gyr after the Big Bang. One avenue of thought is that the SMBHs were able to accrete above the theoretical Eddington limit, at which gravity and radiation pressure balance for ideal spherical symmetry \citep[e.g.,][]{pacucciJWSTCEERSJADES2023,schneiderAreWeSurprised2023,pacucciMildlySuperEddingtonAccretion2024,maiolinoJWSTMeetsChandra2025a}. The Eddington luminosity, $\mathrm{L_{Edd}}$, is related to the SMBH mass by,
\begin{equation}
    \mathrm{L_{Edd} \approx 1.26 \times 10^{38} \, M_{\bullet}},
    \label{eq:ledd}
\end{equation}
and the Eddingtion ratio is,
\begin{equation}
    \label{eq:lambda_edd}
    \mathrm{\lambda_{Edd} = \frac{L_{AGN}}{L_{Edd}}}.
\end{equation}
Clustering analysis of luminous AGN at $z\gtrsim 6$ has shown that the duty cycle of AGN is short with $f_{\mathrm{duty}} \ll 1$ and therefore suggestive of episodic bursts of super-Eddington, $\mathrm{\lambda_{Edd} \gg 1}$, accretion \citep{eilersEIGERVICorrelation2024}. 

Additionally, the initial black hole mass distribution could have been top heavy. Massive stars of $\mathrm{ M_{\star} \sim 10 - 100 \, M_{\odot}}$ typically end their lives as stellar remnant black holes with $\mathrm{M_{\bullet} \sim 10 - 100 \, M_{\odot}}$ \citep[e.g.,][]{hegerHowMassiveSingle2003a}, requiring considerable growth, either from accretion or merging, to explain the distribution of $\mathrm{M_{\bullet} \approx 10^{6} - 10^{9} \, M_{\odot}}$ SMBHs \citep[e.g.,][]{volonteriFormationEvolutionMassive2012,haimanFormationFirstMassive2013}. 

Especially in the metal-poor conditions of the ISM in the first $\approx 1$~Gyr after the Big Bang, ineffective cooling provides some support against fragmentation and the formation of stars, possibly allowing for the emergence of DCBHs with $\mathrm{M_{\bullet} \approx 10^{3}-10^{5} \, M_{\odot}}$ \citep[e.g.,][]{schleicherMassiveBlackHole2013,latifBlackHoleFormation2013,latifFormationSupermassiveBlack2016a}. The seed mass distribution of black holes thus likely spans orders of magnitude, depending on the mechanism \citep{siciliaBlackHoleMass2022b,siciliaBlackHoleMass2022c}.

In the DCBH scenario a prevailing Lyman-Werner radiation field would need to have been established by massive stars to disassociate $\mathrm{H}_{2}$ and remove a source of cooling in primordial dark matter haloes \citep{agarwalUbiquitousSeedingSupermassive2012}. Interestingly, some LRDs have been shown to neighbour UV luminous galaxies whose proximity, $\lesssim 0.3 \, \mathrm{kpc}$, could have provided the necessary radiation field to assist DCBH formation \citep{baggenConnectingDotsUVbright2026a,pacucciMoreNumerousBluer2026,golubchikVENUSWhenRed2026}. 

The abundance of SMBHs is, at the very least, likely emergent from some combination of super-Eddington accretion and top-heavy initial seed black hole masses \citep{volonteriOriginsMassiveBlack2021}. To investigate this we consider a simple toy model. 

Equation~\ref{eq:emc2} shows that the BHAR is connected to $\mathrm{L_{AGN}}$, which is itself connected to the Eddington luminosity through Equations~\ref{eq:ledd} and \ref{eq:lambda_edd}. The rate-of-change of the SMBH mass is thus,
\begin{equation}
    \label{eq:smbhgrowth}
    \frac{\mathrm{dM_{\bullet}(t)}}{\mathrm{dt}} = \frac{1-\epsilon}{\epsilon} \, \lambda_{\mathrm{Edd}} \, \frac{1.26 \times 10^{38}}{c^{2}} \, \mathrm{M_{\bullet}},
\end{equation}
where we pick up a factor of $1-\epsilon$ to account for radiative losses, and hence the difference between the BHAR in Equation~\ref{eq:emc2} and the growth of the SMBH. Equation~\ref{eq:smbhgrowth} is a first-order ordinary differential equation and, under the assumption that neither $\epsilon$ or $\mathrm{\lambda_{Edd}}$ change with time, has exact solution,
\begin{equation}
    \label{eq:smbhsoln}
    \mathrm{
    M_{\bullet}(\mathit{t_{U}(z)} ) = M_{0} \exp \left( \frac{1-\epsilon}{\epsilon} \, \lambda_{Edd} \, \frac{\mathit{t_{U}(z)} - t_{0}}{4.5\times10^{8}}\right )
    },
\end{equation}
where $\mathit{t_{U}(z)}$ is the same as in Equation~\ref{eq:quenching} in units of yr, $\mathrm{M_{0}}$ is the seed black hole mass constant of integration at $\mathrm{t \approx t_{0}}$, which we assume occurs at $z = 30$ \citep{volonteriQuasarsZ6Survival2006a}, and $4.5\times 10^{8} \, \mathrm{yr}$ is the Salpeter timescale (before scaling by $\epsilon$ and $\lambda_{\mathrm{Edd}}$). From Equation~\ref{eq:ledd}, we can recast Equation~\ref{eq:smbhsoln} in terms of the $\mathrm{L_{AGN}}$ at the redshift of observation as,
\begin{multline}
    \label{eq:lagnsoln}
    \mathrm{L_{AGN} = \lambda_{Edd} \times 1.26 \times 10^{38}} \times \\ 
    \mathrm{M_{0} \exp \left( \frac{1-\epsilon  }{\epsilon} \, \lambda_{Edd} \, \frac{\mathit{t_{U}(z)} - t_{0}}{4.5\times10^{8}}\right )}.
\end{multline}

Equation~\ref{eq:lagnsoln} can then be inverted to solve,
\begin{multline}
    \label{eq:solnlambda}
    \mathrm{\lambda_{Edd} = \frac{\epsilon}{(1-\epsilon)} \, \frac{4.5\times10^{8}}{\mathit{t_{U}(z)} - t_{0}}} \times \\ 
    W \left( \frac{1-\epsilon}{\epsilon} \, \frac{\mathit{t_{U}(z)} - \mathrm{t_{0}}}{4.5\times10^{8}}  \mathrm{\frac{L_{AGN}}{1.26\times 10^{38} \, M_{0}}} \right),
\end{multline}
where $W$ is the Lambert $W-$function. 

To solve this equation, we drew $10^{5}$ uniform samples of the radiative efficiency, $\epsilon \sim \mathit{U}(0.06, 0.4)$, and the initial logged seed mass, $\log_{10}(\mathrm{M_{0}})\sim\mathit{U}(2,5)$, to sample the theoretical bounds of each quantity. Each of these quantities were sampled independently. Using these samples and the posterior distributions of $\mathrm{L_{AGN}}$ from the SED fits we computed $\lambda_{\mathrm{Edd}}$ and the mass of the SMBH at the redshift of observation. Treating the parameters $\epsilon$ and $\mathrm{M_{0}}$ as uniformly random variables meant that pessimistic uncertainties could be incorporated.

\begin{figure*}
    \centering
    \includegraphics[width=\linewidth]{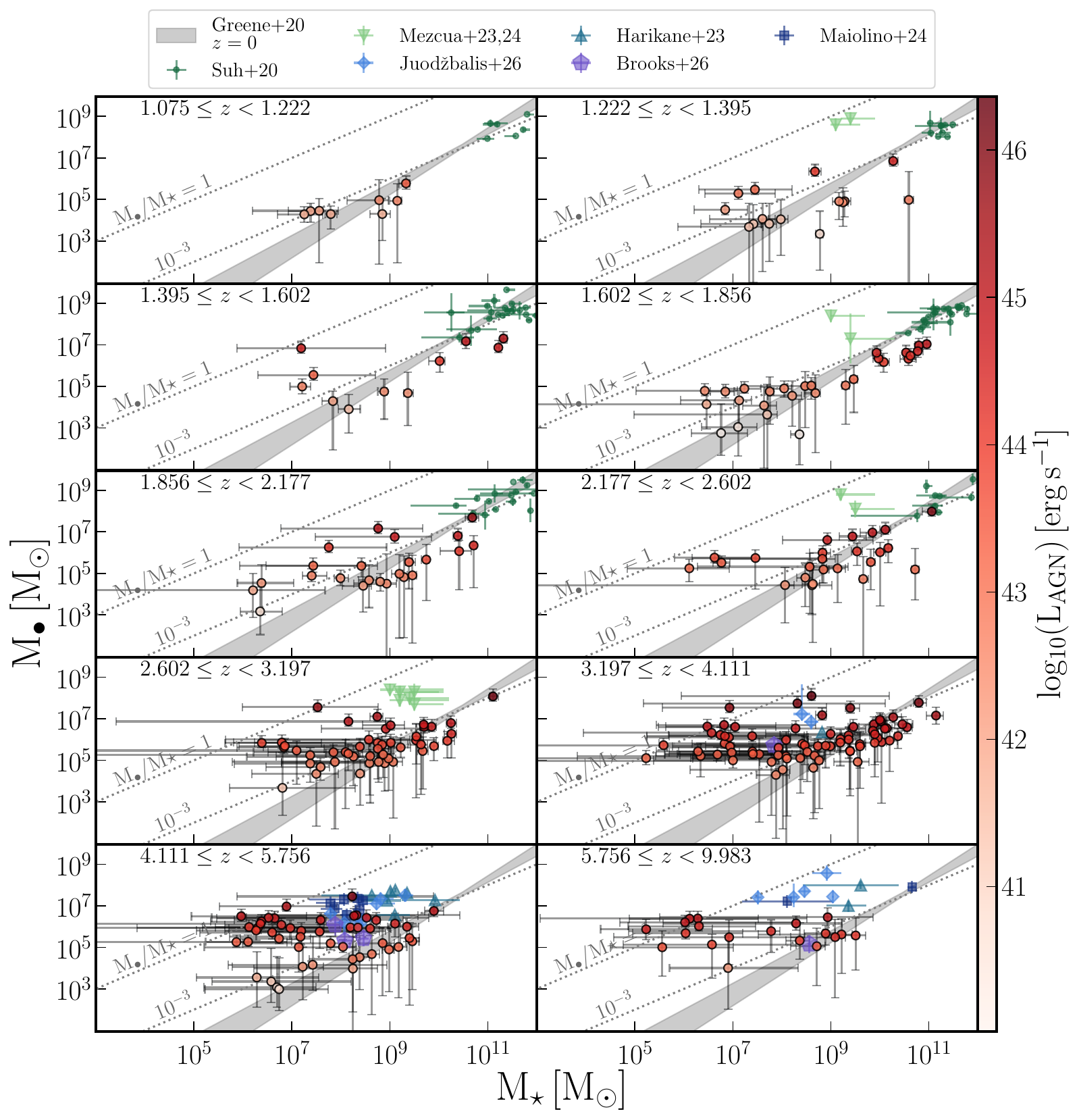}
    \caption{In every panel we show the $\mathrm{M_{\bullet} - M_{\star}}$ relation in each redshift bin as indicated. Error bars on the points show the 1$\sigma$ uncertainties from the posterior distributions. $\mathrm{M_{\bullet}}$ is calculated with our toy model and uncertainties subsume those from $\epsilon$ and $\mathrm{M_{0}}$. The colour of each point is associated with the inferred $\mathrm{L_{AGN}}$ as indicated with the colour bar. Dotted lines show constant ratios of $\mathrm{M_{\bullet}/M_{\star} = 10^{-3}, 1}$. Filled grey lines show the results of the $z=0$ fit from \citet{greeneIntermediateMassBlackHoles2020a}. Literature results from \citet{mezcuaOvermassiveBlackHoles2023,mezcuaOvermassiveBlackHoles2024,harikaneJWSTNIRSpecFirst2023,maiolinoJADESDiversePopulation2024,juodzbalisJADESComprehensiveCensus2026,brooksMonstersMoreComplete2026} are shown as indicated in the figure legend.}
    \label{fig:mstarsmbh}
\end{figure*}

Figure~\ref{fig:mstarsmbh} shows our derived $\mathrm{M_{\star} - M_{\bullet}}$ relation for our sample. 
At $\mathrm{M_{\star} \gtrsim 10^{8} \, M_{\odot}}$ the galaxies tend to lie on the local, $z=0$ relation of \citet{greeneIntermediateMassBlackHoles2020a}, near a stellar mass invariant constant of $\mathrm{M_{\star}/M_{\bullet} \approx 10^{-3}}$. Especially in the redshift bins $z \gtrsim 3$, SMBHs within the galaxies of $\mathrm{M_{\star} \lesssim 10^{7} \, M_{\odot}}$ are $\approx 3$~dex over-massive and close to a mass ratio, $\mathrm{M_{\star}/M_{\bullet} \approx 1}$. 

There is some tension with the $z \gtrsim 3$ results from \citet{maiolinoJADESDiversePopulation2024,juodzbalisJADESComprehensiveCensus2026} who investigate NIRSpec confirmed broad-line sources, with their points lying $\sim 1$~dex higher at $\mathrm{M_{\star} \approx 10^{8} - 10^{9} \, M_{\odot}}$. A similar offset is observed between our results  and those of \citet{mezcuaOvermassiveBlackHoles2023,mezcuaOvermassiveBlackHoles2024} who infer the stellar and SMBH mass for $\mathrm{M_{\star} \approx 10^{9} \, M_{\odot}}$ dwarf galaxies at $z \lesssim 3$ in the VIMOS Public Extragalactic Redshift Survey. Part of this difference between these literature data may therefore be a result of selection for the spectroscopic sample, probing among the most massive SMBHs in comparison to the host galaxy. 

Indeed, we find that our measurements agree well with the results of \citet{brooksMonstersMoreComplete2026} who perform stacking experiments of NIRSpec spectra to obtain representative measurements for the average galaxy, with $\mathrm{M_{\star} \approx 10^{8} \, M_{\odot}}$ and at $z\approx 3-6$. Furthermore, our results are generally in agreement with the trends at $z \lesssim 2.5$ from \citet{suhNoSignificantEvolution2020} obtained for $\mathrm{M_{\star} \approx 10^{10} - 10^{12} \, M_{\odot}}$ galaxies, instead of only dwarf galaxies as done in \citet{mezcuaOvermassiveBlackHoles2023,mezcuaOvermassiveBlackHoles2024} at similar redshifts. Ultimately, SMBHs inhabit galaxies with a range of stellar masses across redshifts, and over-massive SMBHs are not altogether unexpected given the demographics of galaxies hosting AGNs in the Universe.

\subsection{Eddington ratios} \label{sec:eddratios}
\begin{figure*}
    \centering
    \includegraphics[width=\linewidth]{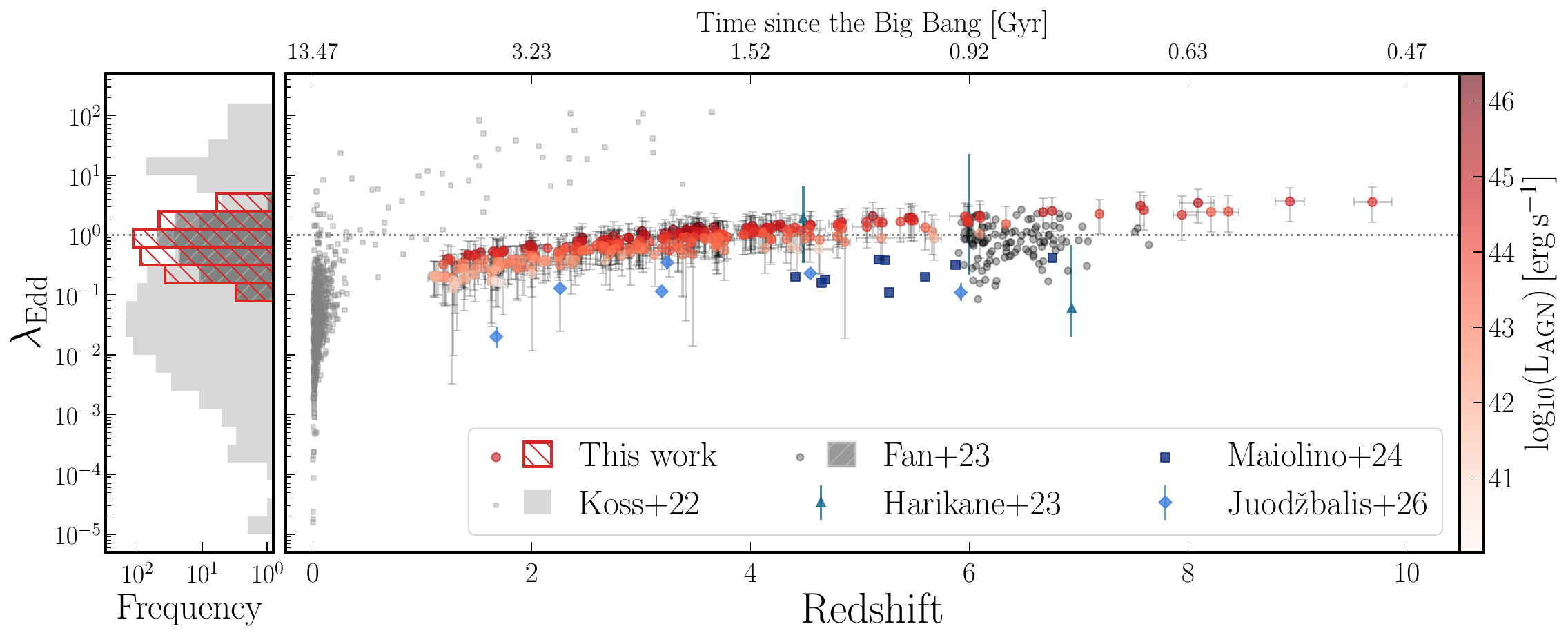}
    \caption{The Eddington ratio against redshift. Error bars show $1\sigma$ uncertainties and points are coloured according to their $\mathrm{L_{AGN}}$ and indicated by the colour bar. Literature results from \citet{kossBASSXXIIBASS2022, fanQuasarsIntergalacticMedium2023, harikaneJWSTNIRSpecFirst2023, maiolinoJADESDiversePopulation2024, juodzbalisJADESComprehensiveCensus2026} are shown as indicated in the figure legend. The left most panel shows the histogram of $\mathrm{\lambda_{Edd}}$ for our results and those of \citet{kossBASSXXIIBASS2022,fanQuasarsIntergalacticMedium2023} as indicated in the legend. The horizontal dotted line shows $\mathrm{\lambda_{Edd}} = 1$.}
    \label{fig:ledd}
\end{figure*}
The Eddington ratios calculated with Equation~\ref{eq:solnlambda} are shown in Figure~\ref{fig:ledd}. The $\mathrm{\lambda_{Edd}}$ values span a range from $\approx 0.1$ to $\approx 3.7$. Compared to $z \lesssim 3$ results from \citet{kossBASSXXIIBASS2022}, our inferred  $\mathrm{\lambda_{Edd}}$ values are skewed higher, with the median difference $\approx 1$~dex that is qualitatively consistent with AGN downsizing. The results from \citet{kossBASSXXIIBASS2022} were obtained from the Swift BAT AGN Spectroscopic Survey (BASS DR2) across the entire sky and hence capture $\mathrm{\lambda_{Edd} \approx 100}$ super-Eddington SMBHs at $z \gtrsim 3$, by virtue of sampling the rarest and most luminous AGNs that are only detectable in such large surveys. 

There is $\sim 1$~dex of tension between our results and those from \citet{maiolinoJADESDiversePopulation2024,juodzbalisJADESComprehensiveCensus2026} that reflects a similar level of disagreement as seen for the SMBH masses in Figure~\ref{fig:mstarsmbh}.

Most of our sample at $z \gtrsim 3$ have super-Eddington accretion rates and vice versa for $z \lesssim 3$. This is consistent with the majority of early SMBH growth having proceeded in super-Eddington bursts of accretion and resulting in low duty cycles \citep{schneiderAreWeSurprised2023, eilersEIGERVICorrelation2024}. Our values at $z \lesssim 6$ agree with the results from \citet{harikaneJWSTNIRSpecFirst2023}. The results from \citet{maiolinoJADESDiversePopulation2024,juodzbalisJADESComprehensiveCensus2026} are nowhere super-Eddington and are representative of fainter AGNs than those from \citet{harikaneJWSTNIRSpecFirst2023}. 

We also compare our results to \citet{fanQuasarsIntergalacticMedium2023} who compiled archival $z>5.9$ spectroscopically confirmed broad-line quasars from the pre-\textit{JWST} literature. They calculate the SMBH masses from the broad component of the Mg II line and the bolometric luminosity from the \SI{3000}{\angstrom} luminosity. Our results are in broad agreement with these observations. This suggests that some of the tension with the more recent NIRSpec results is driven by selection effects. 

\citet{harikaneJWSTNIRSpecFirst2023, maiolinoJADESDiversePopulation2024,juodzbalisJADESComprehensiveCensus2026} infer $\mathrm{\lambda_{Edd}}$ directly from the NIRSpec spectra using virial SMBH mass estimates \citep{greeneEstimatingBlackHole2005, vestergaardDeterminingCentralBlack2006, reinesRelationsCentralBlack2015} and bolometric corrections to the H$\alpha$ line flux \citep{mclureCosmologicalEvolutionQuasar2004,richardsSpectralEnergyDistributions2006, netzerAccretionStarFormation2009, sternMidinfraredSelectionActive2012}. Their results may therefore be considered as instantaneous $\mathrm{\lambda_{Edd}}$ measurements at the epoch of observation. 

The liberal assumption in our computation of $\mathrm{\lambda_{Edd}}$ and $\mathrm{M_{\bullet}}$ is that neither $\mathrm{\lambda_{Edd}}$ nor $\epsilon$ varies with time. This assumption allowed us to smoothly connect the evolution of the black hole mass to the largely instantaneous observation of $\mathrm{L_{AGN}}$. If accretion is episodic, then we have reason to believe that $\mathrm{L_{AGN}}$, $\mathrm{\lambda_{Edd}}$ and $\epsilon$, or some combination thereof, are oscillatory functions of time, in which case the neat general solution in the toy model breaks down. Therefore, we interpret our values of $\mathrm{\lambda_{Edd}}$ to be the time-average of the accretion history since the onset of SMBH growth. 

Our toy model also considers only secular growth of SMBHs within their own systems and may therefore miss a key channel of growth from mergers, especially at $z \gtrsim 4$ \citep[e.g.,][]{duanGalaxyMergersEpoch2026}. Hence, it is important to emphasise that there are methodological differences between our illustrative $\mathrm{\lambda_{Edd}}$ calculations and the spectroscopic results from the literature.

\subsection{Application to LRDs}\label{sec:lrds}
In Figure~\ref{fig:sfr_bhar_evol}, we found that the transition from AGN to stellar dominance occurs at the critical stellar mass of $\mathrm{M_{\star} \approx 10^{7} \, M_{\odot}}$. We draw attention to this value of the stellar mass because of its relation to the SEDs of LRDs. \citet{inayoshiSpectralUniformityLittle2026} suggest that the rest-frame UV radiation in LRDs originates from young, UV luminous stellar populations with $\mathrm{M_{\star} \approx 10^{7} \, M_{\odot}}$. In terms of the evolution, the growth of the stellar mass in LRDs eventually injects feedback that disrupts the SMBH complex and ultimately drives a transition to `normal' AGN phases and, we anticipate, a convergence to stellar dominance at later redshifts as we see in Figure~\ref{fig:bhar_sfr}.

So far, our entire analysis has employed dusty torus models to describe the AGN component in galaxies, including LRDs. We isolate a sample of NIRCam-selected LRDs in our SMILES/MIDIS fields to explore the differences between the standard dusty torus models and the gas enshrouded ones. 

Our `V-shape' selection abides by the following colour criteria (in magnitudes) from \citet{rinaldiWayWeTally2026}: $\mathrm{F150W - F200W < 1.0}$ and $\mathrm{F277W-F444W > 0.5}$. Additional colour criteria are applied to exclude brown dwarf stars also from \citet{rinaldiWayWeTally2026}: $\mathrm{F115W - F200W > -0.5}$ and $\mathrm{F115W-F277W > -0.5}$ and $\mathrm{(F277W-F444W) < 0.4(F115W-F277W)+1.9}$. A final compactness criterion is applied following \citet{kokorevCensusPhotometricallySelected2024}: $f_{\mathrm{F444W}}(0.\arcsec 32) / f_{\mathrm{F444W}}(0.\arcsec 16) \leq 1.7$, which is the ratio of fluxes in the F444W band measured in $0.32\arcsec / 0.16 \arcsec$ radius apertures. 

This yielded a total of 30 LRD candidates, for which we already had stellar and \citet{fritzRevisitingInfraredSpectra2006} AGN fits. Out of the 30 candidates, only 13 of them preferred \textsc{Stellar+AGN} fits with the \citet{fritzRevisitingInfraredSpectra2006} model. Hence, in this section all 30 LRD candidates are refit with AGN included, but narrowing the fitting limits of AGN luminosity to $\mathrm{L_{AGN} \in [10^{40} - 10^{49}] \, erg \, s^{-1}}$ to better encourage significant AGN solutions. 

\begin{figure}
    \centering
    \includegraphics[width=\linewidth]{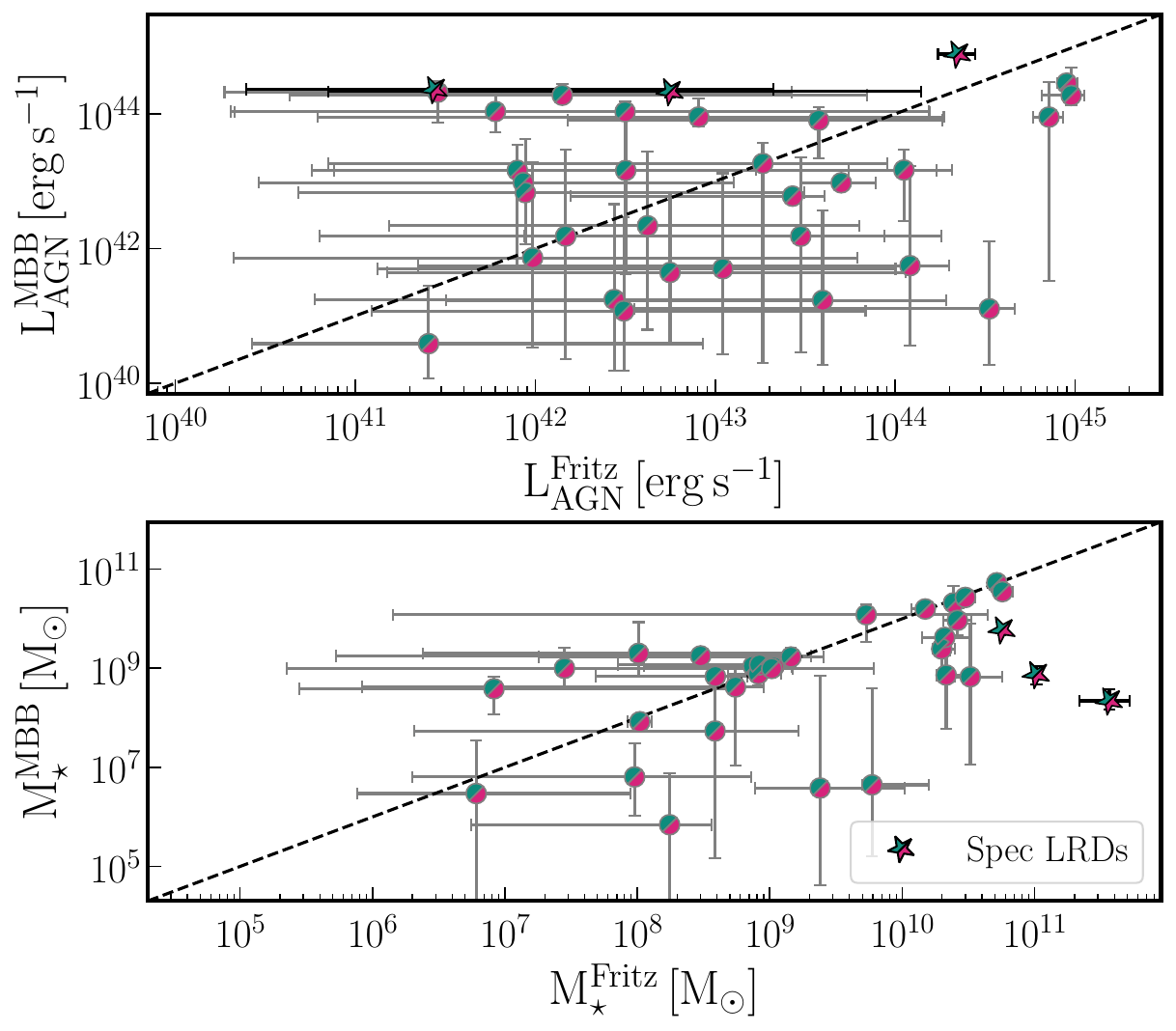}
    \caption{\textit{Top:} comparison of $\mathrm{L_{AGN}}$ between the \citet{fritzRevisitingInfraredSpectra2006} and MBB AGN models. \textit{Bottom:} comparison of $\mathrm{M_{\star}}$ between the \citet{fritzRevisitingInfraredSpectra2006} and MBB AGN models. In all panels, the star symbols show the results for fits to spectroscopic LRDs: MoM-BH$^{*}$-1 \citep{naiduBlackHoleStar2025a}, The Cliff \citep{degraaffRemarkableRubyAbsorption2025} and Rosetta Stone \citep{juodzbalisJADESRosettaStone2024}.}
    \label{fig:lrd_compare}
\end{figure}

The top panel of Figure~\ref{fig:lrd_compare} shows a comparison between the resulting $\mathrm{L_{AGN}}$ between the \citet{fritzRevisitingInfraredSpectra2006} model and the new MBB model. The use of the MBB model tends to lower the bolometric luminosity by a median of $\approx 0.6$~dex because a significant hot dust component is not extrapolated. This could raise the $\mathrm{\frac{SFR_{10}}{BHAR}}$ ratio in Figure~\ref{fig:bhar_sfr} by an average factor of $\approx 4$ for LRDs; though, they are a minority of the sample.

While the uncertainties on $\mathrm{L_{AGN}}$ are large, this result is similar to the findings of \citet{umedaBlackHoleEnvelope2026} who also model the rest-frame optical-IR of LRDs as blackbody functions with Balmer absorption. In their work the rest-frame UV component was modelled with a simple power law. Our integration of the MBB model in \textsc{ProSpect} hence advances spectral modelling of LRDs by leveraging full stellar population synthesis for the host component. The bottom panel of Figure~\ref{fig:lrd_compare} shows a comparison of the resulting stellar mass between the \citet{fritzRevisitingInfraredSpectra2006} and MBB models. The median difference is small, with the MBB model estimating only $\approx 0.06$~dex less stellar mass than \citet{fritzRevisitingInfraredSpectra2006}. The stellar mass can, however, be $\gtrsim 2$~dex lower when the Balmer break has a substantial non-stellar contribution. 

\subsubsection{Beyond the blackbody} \label{sec:beyondblackbody}
Finally, we explore the differences between the standard dusty torus, MBB, quasi photospheric and full cocoon models. For demonstration, we concentrate on three named and spectroscopically confirmed LRDs from the literature: MoM-BH$^{*}$-1 at $z=7.76$ \citep{naiduBlackHoleStar2025a}, The Cliff at $z=3.55$ \citep{degraaffRemarkableRubyAbsorption2025} and Rosetta Stone at $z=2.26$ \citep{juodzbalisJADESRosettaStone2024}. 

We fitted their NIRSpec PRISM spectra that was obtained from version 4.4 of the DAWN \textit{JWST} Archive (DJA)\footnote{\url{https://dawn-cph.github.io/dja/index.html}}. For MoM-BH$^{*}$-1 and The Cliff, MIRI photometry was obtained from the values quoted in their papers. For Rosetta Stone, we included Spitzer channels 3/4 and Multiband Imaging Photometer (MIPS) 24$\mu$m photometry from the CANDELS catalogue \citep{barroCANDELSSHARDSMultiwavelength2019}. 

\begin{figure*}
    \centering
    \includegraphics[width=\linewidth]{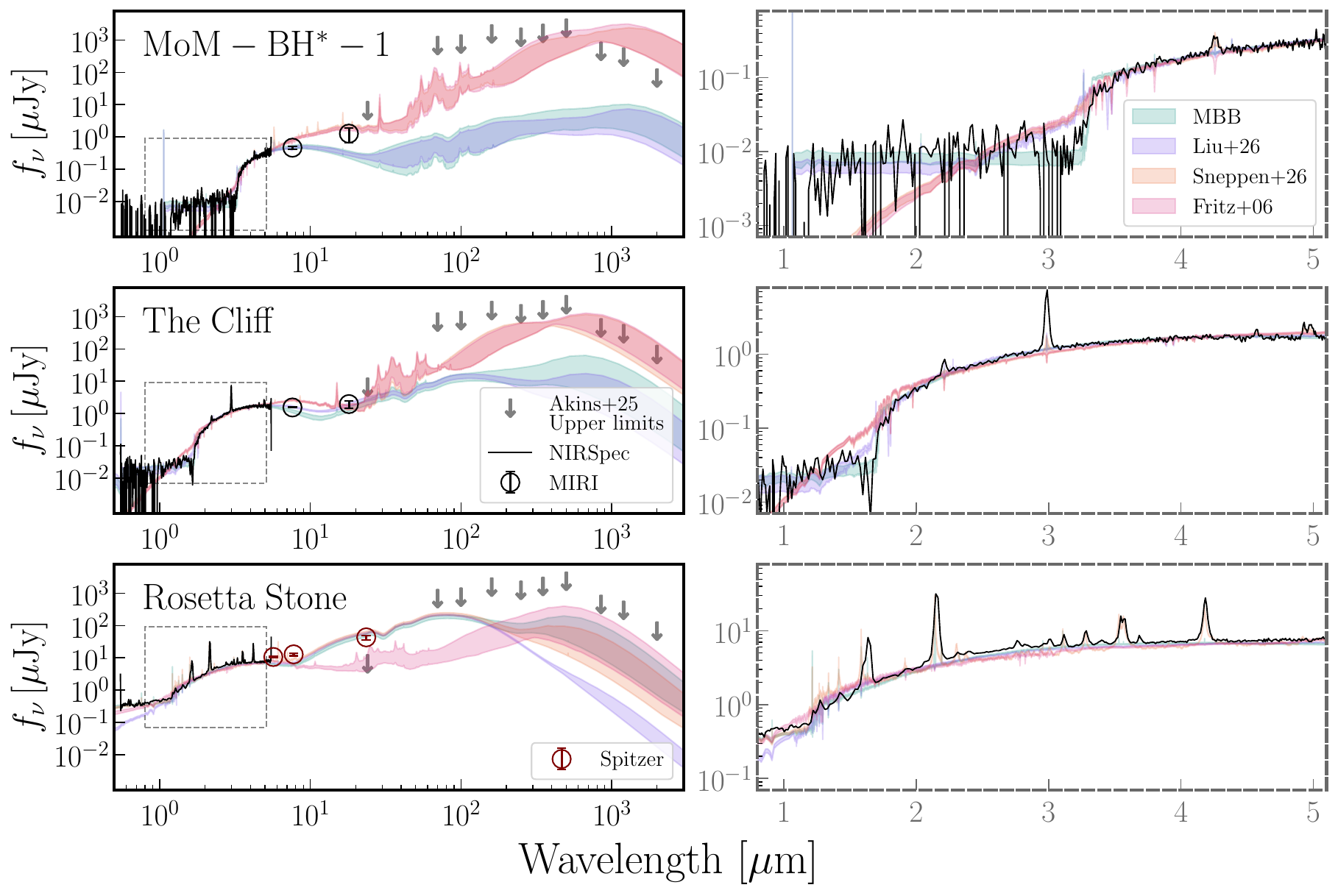}
    \caption{SED fits to LRDs: MoM-BH$^{*}$-1 at $z=7.76$ \citep{naiduBlackHoleStar2025a}, The Cliff at $z=3.55$ \citep{degraaffRemarkableRubyAbsorption2025} and Rosetta Stone at $z=2.26$ \citep{juodzbalisJADESRosettaStone2024}. In every panel, we show the NIRSpec PRISM spectrum and MIRI/Spitzer photometry as indicated in the legends. The filled lines show the 16-\nth{84} percentiles of the four different \textsc{ProSpect} fits from the MBB model, the quasi-photospheric gaseous envelope model \citep{liuSyntheticSpectralLibrary2026}, the gaseous cocoon model \citep{sneppenCocoonComprehensiveExplanation2026} and the standard \citet{fritzRevisitingInfraredSpectra2006} model, as indicated in the legend. The downward facing arrows show the upper limit fluxes from stacks of LRDs \citep[][not used in the SED fitting]{akinsCOSMOSWebOverabundancePhysical2025}. The rightmost panels show a zoom-in of the Balmer break (also highlighted by dashed boxes in the leftmost panels).}
    \label{fig:lrds_seds}
\end{figure*}

The \citet{fritzRevisitingInfraredSpectra2006} models in all three cases tend to prefer subdominant AGN contributions even with the narrowed fitting range of $\mathrm{L_{AGN}}$. This is because the depth of the Balmer break can only be recovered with an old stellar population and there is little room for an additional AGN component. The \textsc{sirocco} models, at least in The Cliff and MoM-BH$^{*}$-1, tend to show similar behaviour as the \citet{fritzRevisitingInfraredSpectra2006} model, preferring an old stellar population to best accommodate the deep Balmer break. The MBB and \citet{liuSyntheticSpectralLibrary2026} models tend to accommodate the deepest non-stellar Balmer breaks. This shows promise that, at least in the most heavily gas-enshrouded LRDs, the emergent SED appears to be well described by the quasi-photospheric models. Expanded model sets of both the cocoon models with \textsc{sirocco} and envelope models with \textsc{tlusty} will be useful for understanding the mechanisms that drive the shape of LRD SEDs.

It is clear that the MIRI/Spitzer observations have some constraining power for discerning between models. The MBB model in particular indicates limited dust emission, below the upper limits of \citet{akinsCOSMOSWebOverabundancePhysical2025}, because a significant dusty torus component is not necessary to explain the spectrum. Hence, we advocate for deep, targeted MIRI observations of LRDs in well studied extragalactic fields. Forthcoming \textit{JWST} cycle 5 programs (ID: 10445 and 12396) will obtain MIRI low-resolution spectroscopy on prominent spectroscopically confirmed LRDs (including those studied in Figure~\ref{fig:lrds_seds}). MIRI imaging redder than the limit of the spectroscopic capabilities ($\lambda_{\mathrm{obs}} \approx 14 \, \mu \mathrm{m}$) will be especially useful for SED fitting constraints.  

\section{Conclusions} \label{sec:conclusions}
We have investigated the assembly of stellar mass and supermassive black holes at $1 \lesssim z \lesssim 10$ using multiwavelength observations of $3218$ galaxies from the \textit{JWST}/MIRI SMILES and MIDIS programs. The main results are summarised below:

\begin{enumerate}
    \item We measured photometry from SMILES and MIDIS using the source extraction software \textsc{ProFound}. \textit{HST}/ACS and \textit{JWST}/NIRCam observations were included from the EPOCHS data set. In total, photometry covering $\mathrm{\lambda_{obs} \approx  0.4 - 25.4\, \mu m}$ was obtained for $4697$ sources. 
    \item Photometric redshifts were computed with \textsc{EAZY} including the `\texttt{agn\_sfhz\_blue}' template set. We find good agreement between our $z_{\mathrm{phot}}$ and $z_{\mathrm{spec}}$ from the literature. 
    \item Using the \textsc{EAZY} redshift posteriors as priors, we performed full SED fitting using the code \textsc{ProSpect} to infer the stellar masses, star formation rates and AGN bolometric luminosities for a sample of $3218$ galaxies at $1 \lesssim z_{\mathrm{phot}} \lesssim 10$. Galaxies that possess a significant AGN were isolated if the contribution of the fitted AGN component to the total SED was above 10 per cent in the rest-frame UV or integrated $3-20 \, \mathrm{\mu m}$. We found that the redshift evolution of the fraction of significant AGNs that occupy galaxies is similar to what has been previously inferred of spectroscopic type-I and type-II populations, although this depends on the adopted selection. A comparison between the inferred AGN luminosities from \textsc{ProSpect} and spectroscopic/X-ray AGN luminosities showed agreement for $\mathrm{L_{AGN} \gtrsim 10^{44} \, erg \, s^{-1}}$, corresponding to $\mathrm{L_{AGN}/M_{\star} \gtrsim 10^{34} \, erg \, s^{-1} \, M_{\odot}}$ in AGN luminosity per unit stellar mass. 
    \item At $z \gtrsim 4$, AGNs were prevalent in $\mathrm{M_{\star} \lesssim 10^{9} \, M_{\odot}}$ galaxies. This could either indicate that we have missed a significant host component from the photometry or that the SMBH would have preferentially grown rather than the stars at this epoch. Black hole accretion rates are greater than star formation rates in galaxies with a critical stellar mass of $\mathrm{M_{\star} \approx 10^{7} \, M_{\odot}}$. AGNs transitioned to lower luminosities and more and more galaxies became quenched from $z \approx 4$ to $z \approx 1$. This is consistent with AGN feedback having inhibited the consumption of the gas supply in galaxies.
    \item We employed a simple toy model derived from accretion physics to compute SMBH masses and Eddington ratios purely from AGN bolometric luminosities. Galaxies with $\mathrm{M_{\star} \gtrsim 10^{8} \, M_{\odot}}$ tend to lie on $z=0$ local relations of the $\mathrm{M_{\star} - M_{\bullet}}$. Especially at $z \gtrsim 3$, galaxies with $\mathrm{M_{\star} \lesssim 10^{7} \, M_{\odot}}$ could have contained $\approx 3$~dex overmassive SMBHs compared to their stellar mass, approaching $\mathrm{M_{\star}/M_{\bullet} \approx 1}$. The inferred time-averaged Eddington ratios are $\lambda_{\mathrm{Edd}} \approx 0.1-3.7$, illustrative of rapid SMBH assembly, particularly at $z \gtrsim 3$ where most AGNs appear to be super-Eddington.
    \item We investigated differences between standard dusty torus AGN models and gas enshrouded models for LRDs. Gas enshrouded models infer a median of $\approx 0.6$~dex lower AGN luminosities than standard dusty models. The stellar mass inferred from the gas enshrouded models can be $\gtrsim 2$~dex lower than dusty torus models because the Balmer break has a substantial non-stellar contribution. We advocate for deep MIRI observations of LRDs to better differentiate between AGN SED models and to stress-test our conclusions that were based on standard dusty torus models. 
\end{enumerate}

\section*{Acknowledgements}
We acknowledge support from the ERC Advanced Investigator Grant EPOCHS (788113). This work is based on observations made with the NASA/ESA \textit{Hubble Space Telescope} (HST) and NASA/ESA/CSA \textit{James Webb Space Telescope} (JWST) obtained from the \texttt{Mikulski Archive for Space Telescopes} (\texttt{MAST}) at the \textit{Space Telescope Science Institute} (STScI), which is operated by the Association of Universities for Research in Astronomy, Inc., under NASA contract NAS 5-03127 for JWST, and NAS 5–26555 for HST. The authors thank all involved with the construction and operation of JWST, without whom this work would not be possible. Some of the data products presented herein were retrieved from the Dawn JWST Archive (DJA). DJA is an initiative of the Cosmic Dawn Center (DAWN), which is funded by the Danish National Research Foundation under grant DNRF140. The authors would like to acknowledge the assistance given by Research IT and the use of the Computational Shared Facility at The University of Manchester. This work was supported by resources provided by the Pawsey Supercomputing Research Centre’s Setonix Supercomputer (\url{https://doi.org/10.48569/18sb-8s43}), with funding from the Australian Government and the Government of Western Australia. Claude Sonnet 5.5 by Anthropic was used to assist with the language, presentation of figures and software development for the Bayesian fitting discussed in Section~\ref{sec:growthsmbhs}. All scientific choices, analyses and presentation were reviewed and verified by the authors.

\section*{Data Availability}
SMILES data can be fround here: \url{https://archive.stsci.edu/hlsp/smiles}. MIDIS data can be found here:  
\url{https://zenodo.org/records/15624625}. Analysis scripts for this paper can be found on \textsc{GitHub:} \url{https://github.com/JordanDSilva/miri-manchester}. Catalogues will be made available upon reasonable request to the corresponding author.  



\bibliographystyle{mnras}
\bibliography{ref, duncan_bib} 




\appendix


\bsp	
\label{lastpage}
\end{document}